\documentstyle[12pt,epsfig]{article}

\begin{document}

\renewcommand{\thefootnote}{\fnsymbol{footnote}}

\newpage
\setcounter{page}{0}

\begin{titlepage}
\begin{flushright}
\hfill{YUMS 97-19}\\
\hfill{KEK-TH-526}\\
\hfill{DESY 97-131}\\
\hfill{August 13, 1997}
\end{flushright}
\vspace{1.0cm}

\begin{center}
{\Large\bf CP Violation at Electron-Positron Colliders\footnote{To appear
           in the Proceedings of the KEK meetings on `$CP$ violation and
           its origin' (1993-1997).} }\\
\hfill{}
\vskip 1.1cm
{M.S.~Baek and S.Y.~Choi} \\
\vskip 0.2cm
{\sl Department of Physics, Yonsei University, Seoul 120-749, Korea}\\
\vskip 0.2cm
{and}\\
\vskip 0.2cm
{K.~Hagiwara}\\
\vskip 0.2cm
{\sl Theory Group, KEK, 1-1 Oho, Tskuba, Ibaraki 305, Japan}
\end{center}

\vskip 1.2cm
\setcounter{footnote}{0}
\begin{abstract}
A detailed, model-independent study of $CP$ violation at present
and future electron-positron colliders is reported. 
Firstly, we investigate $CP$ violation
effects in  $W$ boson and $t$ quark pair production in $e^+e^-$ annihilation
and in two-photon fusion at a next $e^+e^-$ linear collider,
where Compton-backscattered laser light off the electrons
or positrons are employed as a powerful
polarized photon source. Secondly, noting that there do not exist
any direct measurements for the tau-lepton electric dipole moment (EDM),
we address the importance of performing its direct measurements
at $e^+e^-$ collisions off the $Z$-boson pole at TRISTAN, LEPII
and CLEOII. We present a rough comparison of the 
potential of these experiments in the $\tau$ EDM measurements. 
Finally, we report on our recent works for probing $CP$ violation 
in the semileptonic decays of the tau lepton,
which involve two different intermediate resonances with large ratios of 
widths to masses, and which can be most efficiently identified at the 
planned $B$ and proposed $\tau$-charm factories. 
All the $CP$-violation phenomena in the processes under 
consideration, if discovered, imply new $CP$-violation mechanisms 
completely different from the Standard Model $CP$ violation
mechanism through the complex phase of the Cabibbo-Kobayashi-Maskawa  
matrix.
\end{abstract}

\end{titlepage}

\newpage
\renewcommand{\thefootnote}{\alph{footnote}}

\section{Introduction}
\label{sec:Introduction}

The Standard Model (SM) of strong and electroweak interactions\cite{Glashow} 
is widely believed to be the appropriate effective theory below the TeV 
energy scale. There have been many efforts in testing the SM through 
precise measurements over a broad range of energies, but there are no 
confirmed experimental data that disagree with the SM up to now. 
However, it is believed that the SM is not the final theory of Nature
because of several unresolved problems within this model. 
The SM has too many (at least nineteen) free parameters.
This model cannot say anything about the number of families and the origin 
of Higgs mechanism, which breaks the ${\rm SU(2)_L\times U(1)_Y}$ 
electroweak symmetry and is responsible for the masses of weak gauge 
bosons and fermions. 
In addition, it does not include gravitational interactions. 
These motivate a lot of theoretical attempts to understand the SM as a 
low energy effective theory of a more fundamental theory and several  
experimental plans to find new phenomena at and beyond the weak scale.

According to the effective field theory\cite{EFT}, effects of 
more fundamental physics appear as higher dimensional, non-renormalizable 
operators at the SM scale ($v = 250$ GeV), which are usually suppressed by 
powers of the characteristic mass ratio between the SM scale and the large 
mass scale of the fundamental theory.
The most efficient way to find the effects of such higher dimensional 
operators is to investigate the processes or observables which are
suppressed in the SM. In this case, dominant contributions can originate 
from non-renormalizable higher dimensional operators, 
usually from those with the lowest possible dimension.
In this regard, $CP$-violating processes or observables deserve special 
attention since the SM contributions to $CP$-violating processes are often 
very small.

Ever since $CP$ violation was discovered in the neutral Kaon system 
in 1964\cite{Christenson}, this system has been the only place 
where $CP$ violation has been observed.
The neutral Kaon system $CP$ violation can be understood 
within the SM, where the $CP$-violating effects originate from the 
complex phase of the CKM matrix\cite{Cabibbo,Kobayashi}.
Since this phenomenon can also be due to an entirely different source such 
as a superweak mixing of $K^0$ and $\bar{K}^0$\cite{Wolfenstein},
there is no complete understanding of the source of the $CP$ violation 
up to now. 

There is another reason to consider $CP$ violation seriously. 
It is required to explain the baryon asymmetry of the universe\cite{Kolb}. 
No antimatter is observed in our solar system.  Small amount of anti-protons 
that are observed in high energy cosmic rays are consistent with being 
secondary products. These are evidence of the baryon asymmetry 
on the galactic scale.
To generate the baryon asymmetry of the universe starting from the
symmetric universe, three conditions, first proposed by 
Sakharov\cite{Sakharov}, must be satisfied:
B (baryon number) violation, $C$ and $CP$ violation,
and departure from equilibrium. The need of $C$ and $CP$ violation for 
the baryogenesis is obvious when we consider the fact that the baryon number 
is odd under $C$ and $CP$ transformations.
It is shown in recent works\cite{Huet} that the SM 
$CP$-violation mechanism is too insufficient to explain the observed baryon 
asymmetry of the universe.

It is believed that new paradigm of physics should emerge from the 
more fundamental theory when experiments are able to probe energies of the 
electroweak scale. The new theory should explain the cosmological baryon 
asymmetry and may provide us with richer possibilities for $CP$ violation. 
In this aspect, the study of $CP$-violating effects can be an efficient probe
of physics beyond the SM.

As an example, the electric dipole moments (EDM) of quarks and leptons are
ideal from an experimental point of view for
searching evidence of new, non-KM $CP$ violation effects since the KM 
mechanism itself tends to give negligibly small contributions to such 
flavor-conserving processes.  The SM contribution to 
the EDM comes from loop corrections. Several works\cite{Shabalin} 
have shown that the EDM of both the neutron and the electron must vanish 
at the two-loop level. The SM contribution to the EDM appears only at 
the three-loop level, and hence it is extremely small. 
For example, the EDM of the electron arising
from the CKM mechanism has been estimated to be about 
$10^{-38} {\rm e} cm$\cite{Hoogeveen}, which is twelve orders of magnitude 
below the current limit.
 
On the other hand, a number of the most plausible and well-motivated 
extensions of the SM lead to much larger EDM's since these models can give
rise to the EDM's at the one-loop level.
Furthermore, the EDM of a particle is proportional to some positive powers of 
its mass because of the spin flip.
It is, therefore, naturally expected that the top-quark  as 
the most massive quark, the $W$ boson as the most massive 
gauge boson and the $\tau$ lepton as the most massive lepton
are the best candidates to observe non-KM $CP$ violation. 

In the present report, we review a series of our works on a detailed, 
model-independent study of $CP$ violation in processes involving 
the heavy particles, the $W$ boson, the top quark, and the $\tau$ lepton at 
present and future $e^+e^-$ colliders\cite{Choi1,Choi2,Choi3}. 
In section 2, we investigate $CP$ violation effects in  
$W$-boson and $t$-quark pair production in $e^+e^-$ annihilation 
and in two-photon fusion at a next linear $e^+e^-$ collider (NLC)\cite{NLC}, 
where Compton-backscattered laser lights are employed as a powerful polarized photon source\cite{GKS}. 
In section 3, by noting that there do not exist direct measurements of 
the tau-lepton EDM, we address the importance of performing its direct 
measurements at $e^+e^-$ collisions off the $Z$ boson pole at TRISTAN, 
LEPII and CLEOII and present a rough comparison of the potential of 
these experiments in the $\tau$ EDM measurements. Also reported in section 3
is our recent works on probing $CP$ violation in the semileptonic 
$\tau$ decays, which
involve two different intermediate resonances with large ratios of widths
to masses and which can be studied at CLEOII, the planned $B$ factories,
and the proposed $\tau$-charm factories\cite{Gomez}. 
In section 4, we summarize our findings and make a few concluding remarks.

\section{$CP$ Violation at Next $e^+e^-$ Linear Colliders}
\label{sec:NLC}

The possibility of having an NLC  has increased 
significantly in the last few years in parallel with a series
of international workshops called Physics and Experiments with
Linear $e^+e^-$ Colliders\cite{NLC}. There is general consensus for
a 500 GeV NLC  with an integrated luminosity
of the order of 10 fb$^{-1}$ for the first phase. With such experimental 
parameters, NLC produces a copious number of $W$ and top-quark pairs. 
With the expectations that non-KM $CP$ 
violation effects from new interactions beyond the SM are predominant 
in the most massive charged gauge boson and fermion systems, 
the $W$ boson and the $t$ quark, a lot of
works\cite{WbAb,CR,Chang,Atwood,PR,HPZH,GgDs,PkPm,Gounaris,Atwood2} have 
made important contributions to probing $CP$ violation through the production 
and decay processes of these two heavy particles in 
hadron-hadron and $e^+e^-$ collisions.
On the other hand, as one important spin-off of NLC 
highly-energetic photon beams from Compton backscattering of laser
light off electrons or positrons can be utilized to realize electron-photon
and photon-photon collisions. Therefore, recently non-KM $CP$ violation
has been probed in the $W$-pair and $t$-pair production processes
via two-photon fusion\cite{Wudka,Choi2,GbGc}. 

In this section we review a series of our recent works\cite{Choi1,Choi2}
on a detailed, model-independent study of $CP$ violation in $W$-boson 
pair production and top-quark pair production in the $e^+e^-$ and 
two-photon annihilation at an NLC.

The works prior to our works are classified into two categories according to 
their emphasized aspects: (i) the classification of spin and angular 
correlations of the decay products without electron beam 
polarization\cite{WbAb,CR,Chang,HPZH,GgDs,PkPm,KLY} and 
(ii) the use of a few typical $CP$-odd observables with electron beam 
polarization\cite{Atwood,PR}. 
In the first class, they have constructed
a tower of $CP$-odd observables according to their ranks. However, since
the $W$ boson and top quark  are spin-1 and spin-1/2, 
the number of $CP$-odd spin correlations appearing in the processes 
$e^+e^-(\gamma\gamma)\rightarrow W^+W^-$ and 
$e^+e^-(\gamma\gamma)\rightarrow t\bar{t}$ is finite so that 
{\it all the linearly-independent $CP$-odd correlations can be completely
defined}. Then, all the previously-considered correlations are 
expressed in terms of the complete set of linearly-independent 
$CP$-odd correlations. 
In the second class, it has been shown that electron beam polarization
is very crucial for a few specialized $CP$-odd correlations. 
Those works can be easily extended with the complete set of $CP$-odd
correlations in order to investigate which 
$CP$-odd correlations depend crucially on electron polarization and which correlations do not.
     
In order to make our predictions model-independent, we consider all  
possible leading $CP$-violating effects which can be
induced by any extensions of the SM at the weak interaction scale.  
To describe $CP$ violation in the $W$-boson pair production, 
we follow the effective field theory approach with a linear realization 
of the symmetry-breaking sector due to some new interactions which involve
the Higgs sector and the electroweak gauge bosons.
The SU(2)$_L\times$U(1)$_Y$ electroweak gauge symmetry is imposed 
in constructing the higher dimensional effective Lagrangian since there is 
no experimental sign of violation of this symmetry. 
There exist six relevant dimension-six $CP$-odd operators.
On the other hand, the relevant $CP$-odd corrections to the top-quark pair 
production processes are obviously EDM-type
couplings of top and anti-top quarks to a photon and $Z$.

We use two methods in detecting $CP$ violation. One
method makes use of the produced $W$ bosons and top-quarks by measuring
various spin correlations in their final decay products, and the other
method is to employ polarized photon beams to measure various
$CP$-odd polarization asymmetries of the initial states. In the $e^+e^-$ mode,
where the initial $e^+e^-$ state is (almost) $CP$-even due to the very
small electron mass\cite{Hikasa}, only the first method can be used, 
but in the $\gamma\gamma$ mode both methods can be employed. 
We concentrate on the second method in the
two-photon mode by making use of the Compton backscattered laser 
light off the electron or positron beam as a powerful photon source.
The polarization of the scattered high energy photon beams can be controlled 
by adjusting polarizations of initial electron beams and the laser light.
Nevertheless, we will simply consider purely linearly-polarized photon
beams in the present report.

\subsection{Photon Spectrum}
\label{subsec:photon}

Generally, a purely polarized photon beam state is a linear combination
of two helicity states and the photon polarization vector  
can be expressed in terms of two angles $\alpha$ and $\phi$
in a given coordinate system as
\begin{eqnarray}
|\alpha,\phi\rangle =-\cos(\alpha) e^{-i\phi}|+\rangle
           +\sin(\alpha) e^{i\phi}|-\rangle,
\end{eqnarray}
where $0\leq \alpha\leq \pi/2$ and $0\leq \phi\leq 2\pi$.
The degrees of circular and linear polarization are  
$\xi=\cos(2\alpha)$ and $\eta=\sin(2\alpha)$, respectively,
and the direction of maximal linear polarization is denoted
by the azimuthal angle $\phi$. 
For a partially polarized photon
beam it is necessary to re-scale $\xi$ and $\eta$ by its degree
of polarization.

%
\begin{figure}[h]
\hbox to\textwidth{\hss\epsfig{file=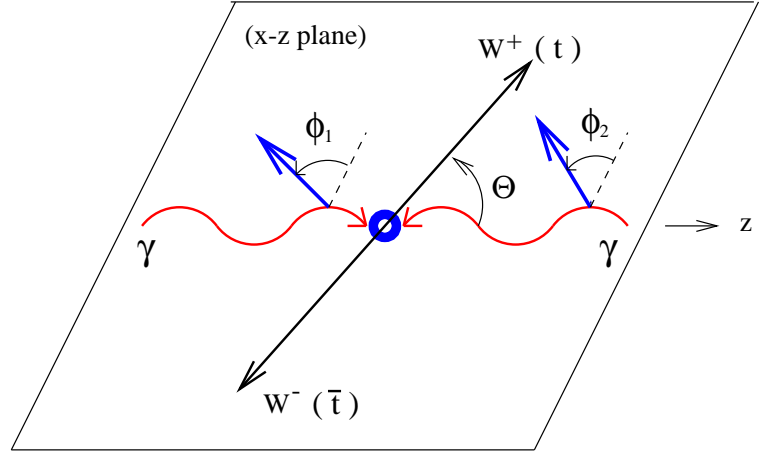,width=10cm}\hss}
\caption{The coordinate system in the colliding $\gamma\gamma$
         c.m. frame. The scattering angle, $\Theta$, and the
         azimuthal angles, $\phi_1$ and $\phi_2$, for the linear
         polarization directions measured from the scattering plane
         are described.}
\label{fig:fig1}
\end{figure}
%

The state vector of the two-photon system in the c.m. frame, 
where one photon momentum is along the positive $z$ direction, is 
\begin{eqnarray}
|\alpha_1,\phi_1;\alpha_2,\phi_2\rangle
       =|\alpha_1,\phi_1\rangle|\alpha_2,-\phi_2\rangle. 
\label{eq:two-photon_wf}
\end{eqnarray}
The angle $\phi_1$ ($\phi_2$) is the azimuthal angle of 
the maximal linear polarization of the photon beam, whose momentum 
is in the positive (negative) $z$ direction, with respect to 
the direction of the $W^+(t)$ momentum in $\gamma\gamma
\rightarrow W^+W^- (\gamma\gamma \rightarrow t\bar{t})$
as shown in Fig.~\ref{fig:fig1}.

Generally, the angular dependence for $\gamma\gamma\rightarrow X$
with two linearly-polarized photon beams, of which the degrees of linear
polarization are $\eta$ and $\bar{\eta}$, is expressed in the form
\begin{eqnarray}
&& {\cal D}(\eta,\bar{\eta};\chi,\phi)
=\Sigma_{\rm unpol}-\frac{1}{2}Re\Bigg[
  \left(\eta{\rm e}^{-i\phi}+\bar{\eta}{\rm e}^{i\phi)}\right)
  {\rm e}^{-i\chi}\Sigma_{02}\Bigg]\nonumber\\
&&\hspace{0.5cm} +\frac{1}{2}Re\Bigg[
  \left(\eta{\rm e}^{-i\phi}-\bar{\eta}{\rm e}^{i\phi}\right)
  {\rm e}^{-i\chi}\Delta_{02}\Bigg]
  +\eta\bar{\eta}Re\Bigg[
  {\rm e}^{-2i\phi}\Sigma_{22}+{\rm e}^{-2i\chi}\Sigma_{00}
  \Bigg],
\label{linear}
\end{eqnarray}
with the subscripts, $0$ and $2$, representing the magnitude of the 
sum of two photon helicities of the initial two-photon system.
Here, two angular variables $\chi$ and $\phi$ are given by
\begin{eqnarray}
\chi=\phi_1-\phi_2,\qquad 
\phi=\phi_1+\phi_2.
\end{eqnarray}
The azimuthal angle difference, $\chi$, is independent of the final state, 
while the azimuthal angle sum, $\phi$, depends on the scattering plane, 
and both angles are invariant with respect to the Lorentz boost along 
the two-photon beam direction. 

\begin{table}[ht]
\caption{$CP$ and $CP\tilde{T}$ properties of the invariant functions 
        and the angular distributions.}
\label{tbl:gamma_symmetry}
\begin{center}
\begin{tabular}{|c|c|c|c|}\hline\hline
 \mbox{ }\hskip 0.2cm $CP$ \mbox{ }\hskip 0.2cm   
&\mbox{ }\hskip 0.2cm $CP\tilde{T}$\mbox{ }\hskip 0.2cm   
&\mbox{ }\hskip 0.2cm Invariant functions \mbox{ }\hskip 0.2cm  
&\mbox{ }\hskip 0.2cm Angular dependences \mbox{ }\hskip 0.2cm  \\
\hline
even & even & $\Sigma_{\rm unpol}$     
            & { } \\
            \cline{3-4}
{ }  & { }  & ${\cal R}(\Sigma_{02})$  
            & $\eta\cos(\phi+\chi)+\bar{\eta}\cos(\phi-\chi)$  \\
            \cline{3-4}
{ }  & { }  & ${\cal R}(\Sigma_{22})$
            & $\eta\bar{\eta}\cos(2\phi)$ \\
            \cline{3-4}
{ }  & { }  & ${\cal R}(\Sigma_{00})$  
            & $\eta\bar{\eta}\cos(2\chi)$ \\ \hline 
even & odd  & ${\cal I}(\Delta_{02})$
            & $\eta\sin(\phi+\chi)+\bar{\eta}\sin(\phi-\chi)$  \\
            \cline{3-4}
{ }  & { }  & ${\cal I}(\Sigma_{22})$  
            & $\eta\bar{\eta}\sin(2\phi)$ \\ \hline
odd  & even & ${\cal I}(\Sigma_{02})$
            & $\eta\sin(\phi+\chi)-\bar{\eta}\sin(\phi-\chi)$  \\
            \cline{3-4}
{ }  & { }  & ${\cal I}(\Sigma_{00})$  
            & $\eta\bar{\eta}\sin(2\chi)$  \\ \hline
odd  & odd  & ${\cal R}(\Delta_{02})$  
            & $\eta\cos(\phi+\chi)-\bar{\eta}\cos(\phi-\chi)$  \\
\hline\hline
\end{tabular}
\end{center}
\end{table}

It is useful to classify the invariant functions $\Sigma$ and
$\Delta$'s, which depend only on the scattering angle $\Theta$,
and the angular distributions according to 
their transformation properties under the discrete symmetries,
$CP$ and $CP\tilde{T}$\footnote{$\tilde{T}$ is the "naive" time
reversal operation which flips particle momenta and spins, but does
not interchange initial and final states}
as shown in Table~\ref{tbl:gamma_symmetry}. 
There exist three $CP$-odd functions; ${\cal I}(\Sigma_{02})$, 
${\cal I}(\Sigma_{00})$ and ${\cal R}(\Delta_{02})$. Here, ${\cal R}$
and ${\cal I}$ stand for real and imaginary parts, respectively.
While the first two terms are $CP\tilde{T}$-even, the last 
term ${\cal R}(\Delta_{02})$ is
$CP\tilde{T}$-odd. Since the $CP\tilde{T}$-odd term 
${\cal R}(\Delta_{02})$ requires an absorptive part in the amplitude,
it is generally expected to be smaller in magnitude than 
the $CP\tilde{T}$-even terms. 
We then can define two $CP$-odd asymmetries from ${\cal I}(\Sigma_{02})$ 
and ${\cal I}(\Sigma_{00})$. 
First, note that $\Sigma_{00}$  is independent of the azimuthal angle 
$\phi$ whereas $\Sigma_{02}$ is not. 
In order to improve the observability we may integrate 
${\cal I}(\Sigma_{02})$ over the azimuthal angle $\phi$ 
with an appropriate weight function. 
Without any loss of generality we can take $\eta=\bar{\eta}$. 
Then, ${\cal I}(\Sigma_{00})$ 
can be separated by taking the difference of the distributions 
at $\chi=\pm\pi/4$  and the ${\cal I}(\Sigma_{02})$ by taking 
the difference of the distributions at $\chi=\pm\pi/2$.
As a result we obtain the following two integrated $CP$-odd
asymmetries\footnote{The authors in Ref.\ \cite{GbGc} have 
also  considered ${\cal I}(\Sigma_{02})$, but not 
   ${\cal I}(\Sigma_{00})$}:
\begin{eqnarray}
\hat{A}_{02}=\left(\frac{2}{\pi}\right)
     \frac{{\cal I}(\Sigma_{02})}{\Sigma_{\rm unpol}},\qquad
\hat{A}_{00}=\frac{{\cal I}(\Sigma_{00})}{\Sigma_{\rm unpol}},
\end{eqnarray}
where the factor $(2/\pi)$ in the $\hat{A}_{02}$ stems from taking 
the average over the azimuthal angle $\phi$ with the weight function
${\rm sign}(\cos\phi)$.

The Compton backscattering process is  
characterized by two parameters $x$ and $y$:
\begin{eqnarray}
x=\frac{4E\omega_0}{m^2_e}
 \approx 15.3\left(\frac{E}{\rm TeV}\right)
             \left(\frac{\omega_0}{\rm eV}\right),\qquad
y=\frac{\omega}{E},
\end{eqnarray}
where $E$ is the electron beam energy and $\omega_0$ the incident
laser beam frequency.
On the average, the backscattered photon energies increase with $x$;
the maximum photon energy fraction is given by
$y_m=x/(1+x)$. Operation below the threshold\cite{GKS} for $e^+e^-$ 
pair production in collisions between the laser beam and the 
Compton-backscattered photon beam requires $x\leq 2(1+\sqrt{2})\approx
4.83$. 

In the two-photon collision case with Compton-backscattered photon beams,
only part of each laser linear polarization is transferred to the 
high-energy photon beam and the degrees of linear polarization 
transfer\footnote{The function $A_{\eta\eta}$ has been 
considered in Ref.~\cite{Kramer}.} are determined by two   
functions, ${\cal A}_\eta$ and ${\cal A}_{\eta\eta}$:  
\begin{eqnarray}
{\cal A}_\eta(\tau)=\frac{\langle \phi_0\phi_3\rangle_\tau}{\langle 
               \phi_0\phi_0\rangle_\tau},\qquad 
{\cal A}_{\eta\eta}(\tau)
     =\frac{\langle \phi_3\phi_3\rangle_\tau}{\langle 
               \phi_0\phi_0\rangle_\tau},
\end{eqnarray}
where $\phi_0(y)=\frac{1}{1-y}+1-y-4r(1-r)$ and $\phi_3(y)=2r^2$
with $r=y/x(1-y)$,  and 
$\tau$ is the ratio of the $\gamma\gamma$ 
c.m. energy squared $\hat{s}$ to the $e^+e^-$ collider energy squared 
$s$. ${\cal A}_\eta$ is for the collision of an 
unpolarized photon beam and a linearly polarized photon beam, and 
${\cal A}_{\eta\eta}$ for that of two linearly 
polarized photon beams.  $\langle \phi_i\phi_j\rangle_\tau$ ($i,j=0,3$) 
is defined as a normalized convolution integral for a fixed value of 
$\tau$. Folding the photon spectrum with the $\gamma\gamma\rightarrow X$
cross section yields two $CP$-odd asymmetries $A_{02}$ and
$A_{00}$, which depend crucially on the two-photon spectrum
and the two linear polarization transfers. 

\begin{figure}[htb]
\hbox to\textwidth{\hss\epsfig{file=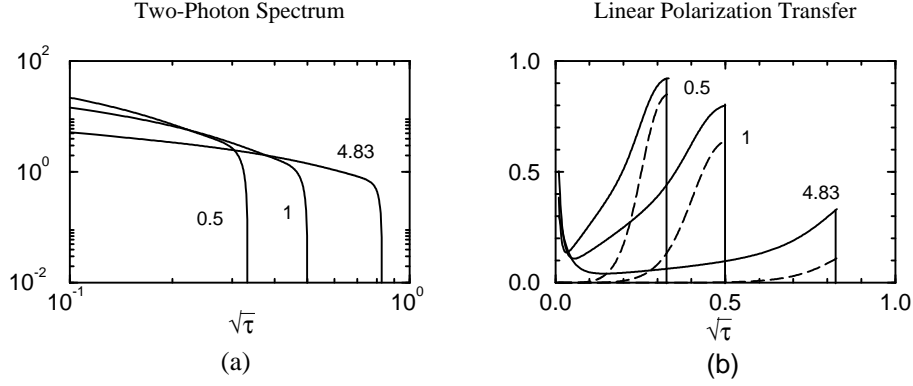,width=12cm}\hss}
\caption{(a) the $\gamma\gamma$ luminosity spectrum and (b) the
          two linear polarization transfers, ${\cal A}_\eta$ (solid lines)
          and ${\cal A}_{\eta\eta}$ (dashed lines), 
          for $x=4E\omega_0/m^2_e=0.5$, $1$ and $4.83$.}
\label{fig:fig3}
\end{figure}

We first investigate the $\sqrt{\tau}$ dependence of the two-photon 
spectrum and the two linear polarization transfers,
${\cal A}_\eta$ and ${\cal A}_{\eta\eta}$ by varying the value of 
the parameter $x$. 
Three values of $x$ are chosen; $x=0.5$, $1$, and $4.83$.
Fig.~\ref{fig:fig3} clearly shows that the energy of two 
photons reaches higher ends for larger $x$ values but the maximum linear 
polarization transfers are larger for smaller $x$ values. 
The parameter $x$ should be kept as large as possible to
reach higher energies. However, larger $CP$-odd asymmetries can be
obtained for smaller $x$ values. Therefore, there should
exist a compromise value of $x$ for the optimal observability of 
$CP$ violation. In this regard, it is very crucial to have a laser with 
adjustable beam frequency.

\subsection{$W$ pair production}
\label{subsec:WW}

To describe $CP$ violation from new interactions among electroweak 
vector bosons in a model-independent way\cite{Model}, we assume that the operators 
with lowest energy dimension dominate the $CP$-odd 
amplitudes and that they respect the electroweak gauge invariance
which is broken spontaneously by an effective SU(2)$_L$-doublet scalar.
Considering $CP$-odd interactions of dimension six composed of electroweak 
gauge bosons and Higgs fields, we can construct an effective Lagrangian
${\cal L}_{\rm eff}$, which is the sum of the SM Lagrangian 
${\cal L}_{\rm SM}$ and the new Lagrangian ${\cal L}_{\rm new}$ defined as
\begin{eqnarray}
{\cal L}_{\rm new}=\frac{1}{\Lambda^2}
           \bigg[f_{B\tilde{B}}{\cal O}_{B\tilde{B}}
               + f_{B\tilde{W}}{\cal O}_{B\tilde{W}}
               + f_{W\tilde{W}}{\cal O}_{W\tilde{W}}
               + f_{\tilde{B}}{\cal O}_{\tilde{B}}
               + f_{\tilde{W}}{\cal O}_{\tilde{W}}
               + f_{WW\tilde{W}}{\cal O}_{WW\tilde{W}}\bigg],
\label{eq:effective}
\end{eqnarray}
where the six $CP$-odd dimension-six operators are as follows 
\begin{eqnarray}
\begin{array}{ll}
  {\cal O}_{B\tilde{B}}=g^{\prime 2}(\Phi^\dagger\Phi)
                        B_{\mu\nu}\tilde{B}^{\mu\nu}, &
  {\cal O}_{B\tilde{W}}=gg^\prime(\Phi^\dagger\sigma^I\Phi)
                        B_{\mu\nu}\tilde{W}^{I\mu\nu}, \\
  {\cal O}_{W\tilde{W}}
   =g^2(\Phi^\dagger\Phi)W^I_{\mu\nu}\tilde{W}^{I\mu\nu}, &
  {\cal O}_{\tilde{B}}
   =ig^\prime\left[(D_\mu\Phi)^\dagger(D_\nu\Phi)\right]
       \tilde{B}^{\mu\nu},\\
  {\cal O}_{\tilde{W}}
   =ig\left[(D_\mu\Phi)^\dagger\sigma^I(D_\nu\Phi)\right]
      \tilde{W}^{I\mu\nu}, &
  {\cal O}_{WW\tilde{W}}=g^3\epsilon^{IJK}\tilde{W}^{I\mu\nu}
                       W_\nu^{J\rho}W^K_{\rho\mu},
\end{array}
\label{eq:dimension-six}
\end{eqnarray}
with the Higgs-doublet $\Phi$,
$\tilde{W}^{I\mu\nu}=\frac{1}{2}\epsilon^{\mu\nu\alpha\beta}
W^I_{\alpha\beta}$, and $\tilde{B}^{\mu\nu}
=\frac{1}{2}\epsilon^{\mu\nu\alpha\beta}
\tilde{B}_{\alpha\beta}$.
Table~\ref{tbl:dimension-six vertices}
shows which vertices already exist in the SM at tree level 
and which  new vertices appear from the new dimension-six $CP$-odd 
operators.
\begin{table}[htb]
\caption{Vertices relevant for the processes $e^+e^- \rightarrow W^+W^-$ and
         $\gamma\gamma\rightarrow W^+W^-$ in the effective Lagrangian 
         with the six dimension-six $CP$-odd operators.}
\label{tbl:dimension-six vertices}
\begin{center}
\begin{tabular}{|c|c|c|c|c|c|}\hline\hline
 \mbox{ }\hskip 0.2cm Vertex\mbox{ }\hskip 0.2cm 
&\mbox{ }\hskip 0.2cm $\gamma WW$ \mbox{ }\hskip 0.2cm 
&\mbox{ }\hskip 0.2cm $ZWW$ \mbox{ }\hskip 0.2cm 
&\mbox{ }\hskip 0.2cm $\gamma\gamma WW$ \mbox{ }\hskip 0.2cm 
&\mbox{ }\hskip 0.2cm $HWW$ \mbox{ }\hskip 0.2cm 
&\mbox{ }\hskip 0.2cm $\gamma\gamma H$ \mbox{ }\hskip 0.2cm\\ 
\hline
SM &  {\large o} & {\large o} &  {\large o}  &  {\large o}  & x \\ 
\hline
\mbox{ }\hskip 0.2cm ${\cal O}_{B\tilde{B}}$\mbox{ }\hskip 0.2cm 
   &  x  & x &  x  &  x  &  {\large o}       \\ 
\hline
\mbox{ }\hskip 0.2cm ${\cal O}_{B\tilde{W}}$\mbox{ }\hskip 0.2cm 
   &  {\large o} & {\large o}  &  x  &  x  &  {\large o}       \\ 
\hline
\mbox{ }\hskip 0.2cm ${\cal O}_{W\tilde{W}}$\mbox{ }\hskip 0.2cm
   &  x  & x &  x  &  {\large o}  &  {\large o}       \\ 
\hline
\mbox{ }\hskip 0.2cm ${\cal O}_{\tilde{B}}$\mbox{ }\hskip 0.2cm 
   &  {\large o} & {\large o} &  x  &  x  &  x   \\ 
\hline
\mbox{ }\hskip 0.2cm ${\cal O}_{\tilde{W}}$\mbox{ }\hskip 0.2cm 
   & {\large o} & {\large o}  &  x  &  {\large o}  & x      \\ 
\hline
\mbox{ }\hskip 0.2cm ${\cal O}_{WW\tilde{W}}$\mbox{ }\hskip 0.2cm 
   &  {\large o} & {\large o}  &  {\large o} & x &  x       \\ 
\hline\hline
\end{tabular}
\end{center}
\end{table}

For the sake of an efficient analysis in the following, 
we define four new dimensionless form factors, 
$Y_i$ ($i=1$ to $4$), which are related with the coefficients, 
$f_i$'s ($i=B\tilde{B},B\tilde{W},W\tilde{W},\tilde{B},\tilde{W},WW\tilde{W}$) 
as
\begin{eqnarray}
&&Y_1=\left(\frac{m_W}{\Lambda}\right)^2
      \bigg[f_{B\tilde{W}}
         +\frac{1}{4}f_{\tilde{B}}+f_{\tilde{W}}\bigg],\qquad 
  Y_2=\left(\frac{m_W}{\Lambda}\right)^2
      \frac{g^2}{4}f_{WW\tilde{W}},\nonumber\\
&&Y_3=\left(\frac{m_W}{\Lambda}\right)^2
      \bigg[f_{W\tilde{W}}+\frac{1}{4}f_{\tilde{W}}\bigg],\qquad
  Y_4=\left(\frac{m_W}{\Lambda}\right)^2
      \bigg[f_{B\tilde{B}}-f_{B\tilde{W}}-f_{W\tilde{W}}\bigg].
\end{eqnarray}
We note in passing that if all the coefficients, $f_i$, are of the 
similar size, then $Y_2$ would be about ten times smaller than the 
other form factors in size because of the factor $g^2/4\sim 0.1$.

\subsubsection{Electron-positron mode}

The angular dependence
$\Sigma_{L,R}(\Theta;\theta,\bar{\theta};\phi,\bar{\phi})$
of a sequential process $e^+e^-\rightarrow W^+W^-\rightarrow
(f_1\bar{f}_2)(f_3\bar{f_4})$ can be 
decomposed in terms of eighty-one orthogonal functions 
${\cal D}_\alpha$'s of the $W^+$ and $W^-$ decay products as
\begin{eqnarray}
\Sigma_{L,R}(\Theta;\theta,\bar{\theta};\phi,\bar{\phi})
  =
\sum_{\alpha=1}^{81}{\cal P}_{\alpha L,R}(\Theta)
     {\cal D}_\alpha(\theta,\bar{\theta};\phi,\bar{\phi}),
\label{eq:angular_dependence:W}
\end{eqnarray}
where $\Theta$ is the scattering angle between $e^-$ and $W^-$, 
and $\theta(\bar{\theta})$ and $\phi(\bar{\phi})$ are
the angular variables of the $W^-$ and $W^+$ decay products.
All the terms, ${\cal P}_{\alpha X}$ and ${\cal D}_\alpha$, can 
be divided into four categories under $CP$ and $CP\tilde{T}$: 
even-even, even-odd, odd-even, and odd-odd terms\cite{Choi1}. 
There exist {\it thirty-six} independent $CP$-odd terms among which 
{\it eighteen} terms are $CP\tilde{T}$-even and the other {\it eighteen} 
$CP\tilde{T}$-odd.

Including electron polarization, most of the  
works\cite{HPZH,GgDs,Chang,PkPm} prior to our works
have considered four special $CP$-odd and $CP\tilde{T}$-even asymmetries, 
of which two are 
essentially equivalent to the so-called triple vector products, and four 
new $CP$-odd and $CP\tilde{T}$-odd asymmetries in addition to the two 
conventional lepton energy asymmetries. 
Clearly, the present analysis shows that much more $CP$-odd asymmetries 
are available.

Observables which are constructed from the momenta of the
charged leptons originating from $W^+$ and $W^-$
decay are directly and most easily measurable in future experiments. 
Therefore, the $W$ leptonic decay channels together with the 
corresponding charge-conjugated ones are exclusively
used in the following analysis. 
The first set of observables under consideration 
involves the momentum of a lepton from $W^-$ decay 
correlated with the momentum of a lepton from
$W^+$ decay in the sequential process
\begin{eqnarray}
e^+(\vec{p}_{\bar{e}})+e^-(\vec{p}_{e})
   \rightarrow W^+ + W^-
   \rightarrow l^+(\vec{q}_+)+l^-(\vec{q}_-)+X,
\label{eq:inclusive;eeww}
\end{eqnarray}
As shown in Ref.~\cite{BN} a tower of $CP$-odd observables can be
in principle constructed, 
among which a few typical $CP$-odd observables\cite{BN} are listed
in the following:
\begin{eqnarray}
&& A_1=\hat{p}_e\cdot(\vec{q}_+\times\vec{q}_-),\nonumber\\
&& T_{ij}=(\vec{q}_--\vec{q}_+)_i(\vec{q}_-\times\vec{q}_+)_j
        +(i\leftrightarrow j),\nonumber\\
&& A_E=E_+-E_-,\qquad
   A_2=\hat{p}_e\cdot(\vec{q}_++\vec{q}_-),\nonumber\\
&&Q_{ij}=(\vec{q}_-+\vec{q}_+)_i(\vec{q}_--\vec{q}_+)_j
        +(i\leftrightarrow j)
        -\frac{2}{3}\delta_{ij}(\vec{q}^2_--\vec{q}^2_+).
\label{observable}
\end{eqnarray}
The observables $T$ and $A_1$ are $CP\tilde{T}$-even, whereas 
$Q$, $A_2$ and $A_E$ are $CP\tilde{T}$-odd.
Certainly, all the $CP$-odd observables can be expressed as a linear
combination of a fixed number of linearly-independent observables whose
classification depends only on the spins of the final particles\cite{Choi1}.
Because of the lack of space, we refer to the work\cite{Choi1} for
a more detailed explanation for the point and, following the same procedure 
as Ref.~\cite{BN}, we use the observables (\ref{observable}) in probing 
$CP$ violation in $e^+e^-\rightarrow W^+W^-$ in the present report.

The statistical significance of a given observable $O_X$ is determined by
comparing its expectation $\langle O_X\rangle$ with the expectation 
variance $\langle O^2_{X}\rangle_{\rm SM}$ in the SM where
$\langle O_X\rangle_{SM}$ vanishes.
Quantitatively, an observation of any deviation from the SM expectation
with better than one-standard deviations requires
\begin{eqnarray}
\langle O_{X}\rangle\geq
        \sqrt{\frac{\langle O_{X}^2\rangle_{\rm SM}}{N_{WW}}},\qquad 
N_{WW}=\varepsilon\left[B_{X^+}\bar{B}_{X^-}\right]
        {\cal L}_{ee}\sigma(e^+e^-\rightarrow W^+W^-),
\label{eq:deviation;w}
\end{eqnarray}
where $N_{WW}$ is the number of events,  ${\cal L}_{ee}$ 
is the $e^+e^-$ integrated luminosity, 
$B_{X^\pm}$ are the branching fractions of $W^\pm\rightarrow l^\pm \nu_l$,
and $\varepsilon$ is the detection efficiency, which is assumed to be 
unity.

%
\begin{figure}[thb]
\hbox to\textwidth{\hss\epsfig{file=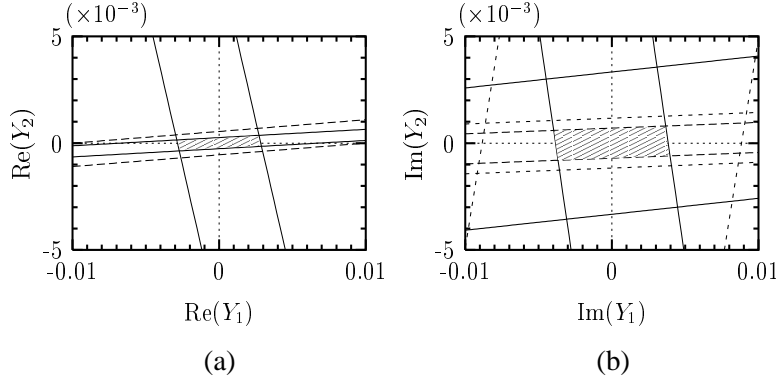,height=5.0cm}\hss}
\caption{(a) The 1-$\sigma$ allowed regions of 
             $Re(Y_1)$ and $Re(Y_2)$ through $A^l_1$ (solid) and 
             $T^l_{33}$ (long-dashed) with polarized electron beams 
             and with the $e^+e^-$ integrated luminosity 10 fb$^{-1}$ 
             at $\sqrt{s}=500$ GeV.  
         (b) The 1-$\sigma$ allowed regions of 
             $Im(Y_1)$ and $Im(Y_2)$ through $A^l_E$ (solid), 
             $A^l_2$ (long-dashed) and $Q^l_{33}$ (dashed) with 
             polarized electron beams and with the $e^+e^-$
             integrated luminosity 10 fb$^{-1}$ at $\sqrt{s}=500$ GeV.}
\label{fig:ylimit}
\end{figure}
%

In determining the 1-$\sigma$ allowed region of the $(Y_1,Y_2)$ plane 
we take the following set of experimental parameters:
\begin{eqnarray}
\sqrt{s}=0.5\ \ {\rm TeV},\qquad L_{ee}=10\ \ {\rm fb}^{-1}.
\label{exp_parameters}
\end{eqnarray}
We note in Fig.~\ref{fig:ylimit} that the use of longitudinal 
electron beam polarization obviates the need for the simultaneous 
measurement of more than one distribution 
and it can greatly enhance the sensitivities to the $CP$-odd parameters
by using two or more $CP$-odd observables.
The 1-$\sigma$ optimal sensitivities to ($Re(Y_1)$, $Re(Y_2)$) and 
($Im(Y_1)$, $Im(Y_2)$) read 
\begin{eqnarray}
&& |Re(Y_1)|\leq 2.9\times 10^{-3},\qquad 
   |Re(Y_2)|\leq 3.7\times 10^{-4},\nonumber\\
&& |Im(Y_1)|\leq 3.9\times 10^{-3},\qquad 
   |Im(Y_2)|\leq 1.0\times 10^{-3}.
\label{optimal bound on real Y_1 and Y_2}
\end{eqnarray}
%

\subsubsection{Two-photon mode}
\label{subsubsec:WW_PP}

In counting experiments where final $W$ polarizations are not
analyzed, we measure only the distributions summed over the final
$W$ polarizations, from which the explicit form of $\Sigma_{\rm unpol}$, 
$ \Sigma_{02}$, $\Delta_{02}$, $\Sigma_{22}$, and  $\Sigma_{00}$
in Eq.~(\ref{linear}) can be obtained.
First of all, we emphasize that ${\cal I}(\Sigma_{00})$ does not require
any identification of the scattering plane as mentioned before. 
Even if one excludes the $\tau^+\tau^-+\not\!{p}$ modes of 1\%, 
the remaining 99\% of the events can be used
to measure ${\cal I}(\Sigma_{00})$. On the other hand, the scattering
plane should be identified to measure ${\cal I}(\Sigma_{02})$. 
Nevertheless, it is worth noting that the charge of the decaying 
$W$ is not needed to extract ${\cal I}(\Sigma_{02})$. 
Therefore, all the modes except for the $l^+l^-+\not\!{p}$ modes 
(9\%) can be used for ${\cal I}(\Sigma_{02})$.

The $\gamma\gamma\rightarrow W^+W^-$ reaction has a much 
larger cross section than heavy fermion-pair production such as 
$\gamma\gamma\rightarrow t\bar{t}$ and, furthermore, the total cross 
section approaches a constant value at high energies. 
At $\sqrt{\hat{s}}=500$ GeV the total cross section is about 80~pb, 
while the $t\bar{t}$ cross section is about 1~pb. So, there exist no
severe background problems. 
In the following analysis we simply assume that all the $W$ pair events 
can be used. 

We present our numerical results for the following set of experimental
parameters:             
\begin{eqnarray}
\sqrt{s}=0.5\ \ {\rm and}\ \  1.0\ \ {\rm TeV},\qquad 
\kappa^2 L_{ee}=20\ \ {\rm fb}^{-1}.
\label{rr_exp}
\end{eqnarray}
Here, $\kappa$ is the $e$-$\gamma$ conversion coefficient in the Compton
backscattering. The parameter $x$, which is dependent on the laser 
frequency $\omega_0$, is treated as an adjustable parameter. 
Folding the photon luminosity spectrum and integrating the 
distributions over the final kinematic variables, we obtain the 
$x$-dependence of available event rates.

Separating the $CP$-odd asymmetries $A_a$ into three parts as
\begin{eqnarray}
A_a=R(Y_1)A^{Y_1}_a+R(Y_2)A^{Y_2}_a+R(Y_4)A^{Y_4}_a,
\end{eqnarray}
and considering each form factor separately,
we then obtain the $1$-$\sigma$ allowed upper bounds of the form factors
($i=1,2,4$)
\begin{eqnarray}
{\rm Max}(|R(Y_i)|_a)
   =\frac{\sqrt{2}}{|A^{Y_i}_a\sqrt{\varepsilon N_{\rm unpol}}|}, 
\end{eqnarray}
if no asymmetry is found.
Here, $\varepsilon$ is for the sum of $W$ branching fractions available, 
which is taken to be $100\%$ for $A_{00}$ and $91\%$ for $A_{02}$.

%
\begin{figure}[ht]
\hbox to\textwidth{\hss\epsfig{file=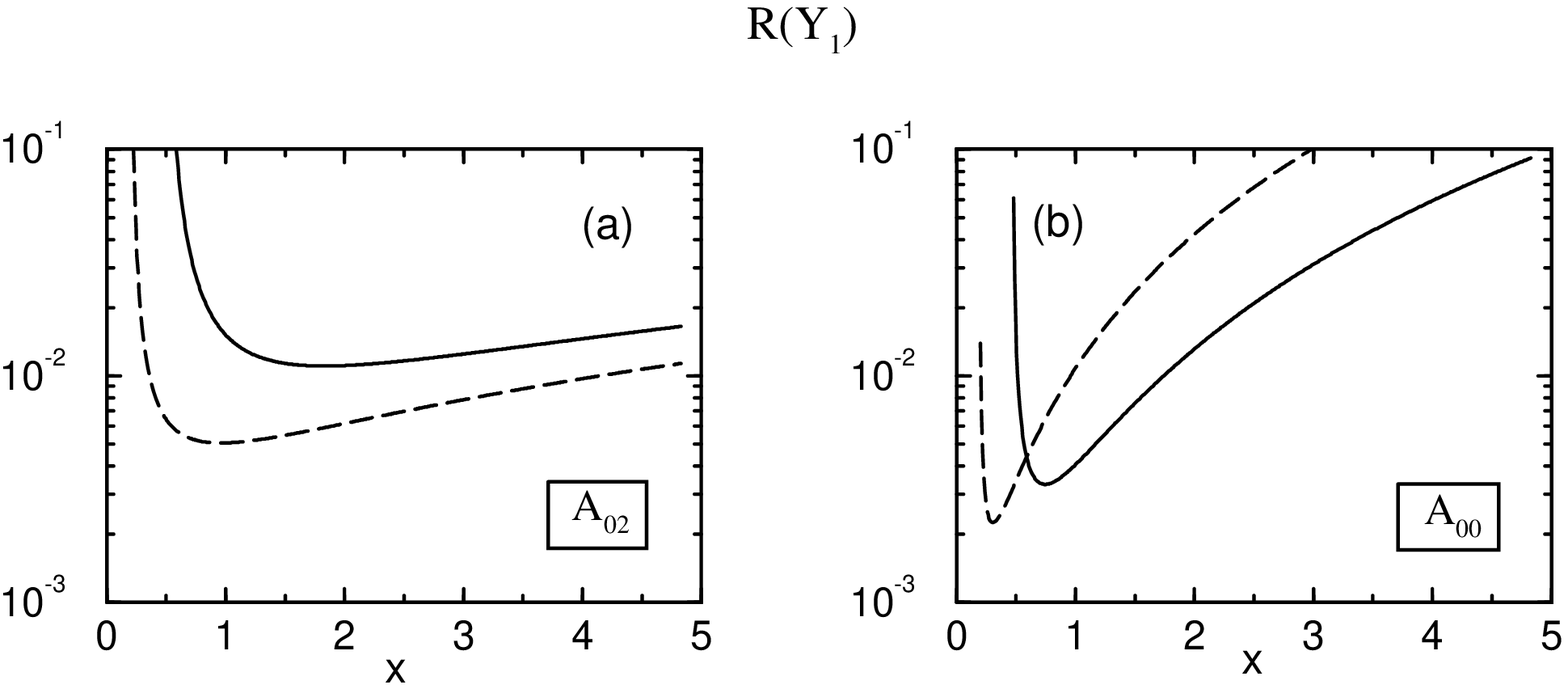,height=3.5cm,width=8cm}\hss}
\vskip 0.3cm
\hbox to\textwidth{\hss\epsfig{file=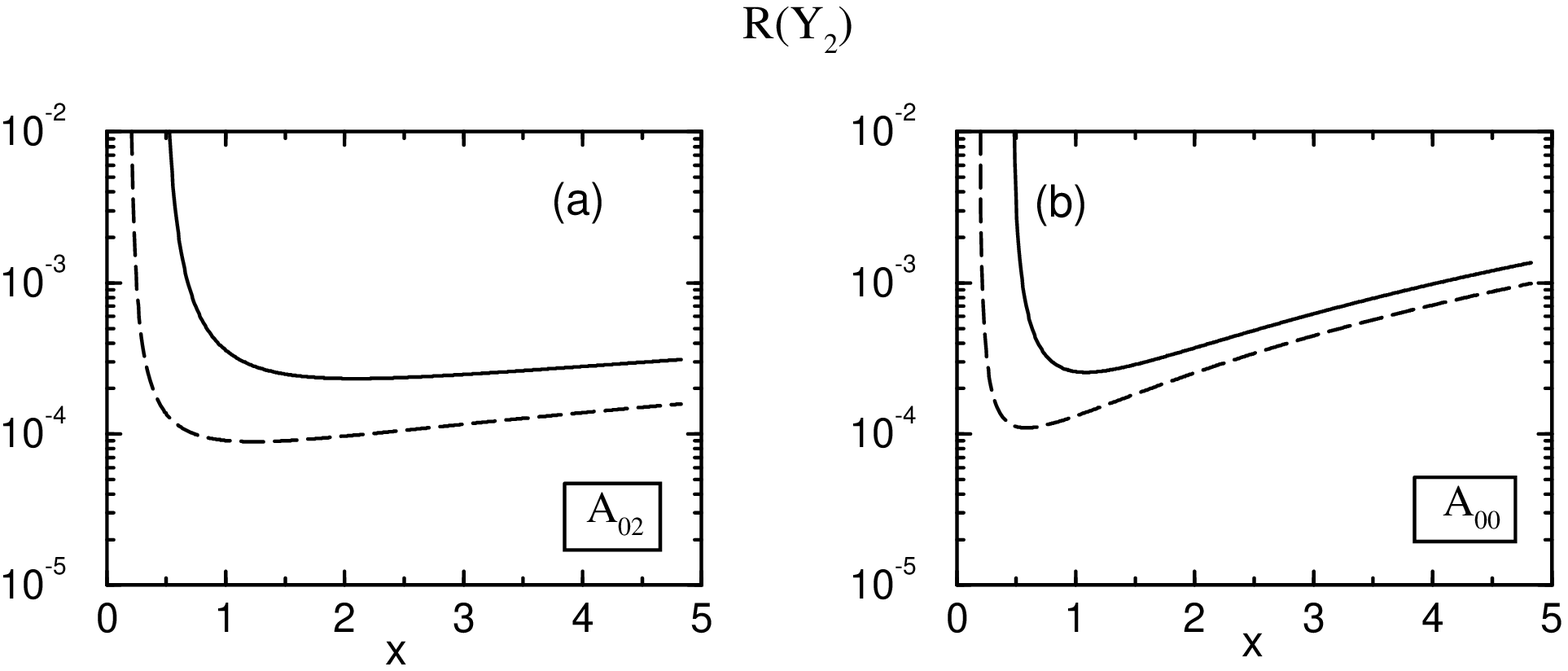,height=3.5cm,width=8cm}\hss}
\caption{The $x$ dependence of ${\rm Max}(|R(Y_1)|)$ and 
         ${\rm Max}(|R(Y_2)|)$ at $\sqrt{s}=0.5$ (solid) and 
         $1.0$ TeV (long-dashed) from (a) $A_{02}$ and 
         (b) $A_{00}$.}
\label{fig:fig4}
\end{figure}

%
Postponing the analysis of the $CP$-violation effects due to 
$Y_4$\cite{Choi1,Gounaris}, we present the analysis of the constraints on $Y_1$
and $Y_2$. Figs.~\ref{fig:fig4}(a) and (b) show the $x$ dependence of 
the 1-$\sigma$ sensitivities 
to $R(Y_1)$ and $R(Y_2)$, which are obtained from $A_{02}$ and $A_{00}$, 
respectively, for $\sqrt{s}=0.5$ TeV (solid) and $\sqrt{s}=1$ TeV 
(long-dashed). 
These figures and Table~\ref{tbl:1-sigma bounds;Y_1 and Y_2} clearly 
exhibit that
(i) the sensitivities, especially through $A_{00}$,
depend strongly on $x$, and  (ii) the optimal sensitivities on $R(Y_2)$ 
are very much improved as $\sqrt{s}$ increases from $0.5$ TeV to 
1 TeV, while those of $R(Y_1)$ are a little improved.
\begin{table}[hb]
\caption{The best $1$-$\sigma$ sensitivities to $R(Y_1)$ 
and $R(Y_2)$, and their corresponding $x$ values for $\sqrt{s}=0.5$ 
and $1$ TeV.}
\label{tbl:1-sigma bounds;Y_1 and Y_2}
\begin{center}
\begin{tabular}{|c|c|c|c|c|}\hline\hline
 ${\rm Asymmetry} $   &  \multicolumn{2}{c|}{$A_{02}$}  
                      &  \multicolumn{2}{c|}{$A_{00}$}  \\
\hline
    $\sqrt{s}$ (TeV)  & $0.5$  &  $1.0$  
                      & $0.5$  &  $1.0$ \\ \hline
       $x$            & $1.83$ &  $0.96$
                      & $0.75$ &  $0.31$\\ \hline
  ${\rm Max}(|R(Y_1)|)$  
      &\mbox{ }\hskip 0.2cm  $1.1\times 10^{-2}$\mbox{ }\hskip 0.2cm  
      &\mbox{ }\hskip 0.2cm  $5.0\times 10^{-3}$\mbox{ }\hskip 0.2cm 
      &\mbox{ }\hskip 0.2cm  $3.2\times 10^{-3}$\mbox{ }\hskip 0.2cm  
      &\mbox{ }\hskip 0.2cm  $2.2\times 10^{-3}$\mbox{ }\hskip 0.2cm \\ \hline
       $x$            & $2.09$ &  $1.23$
                      & $1.11$ &  $0.59$\\ \hline
  ${\rm Max}(|R(Y_2)|)$  
      &\mbox{ }\hskip 0.2cm $2.4\times 10^{-4}$\mbox{ }\hskip 0.2cm 
      &\mbox{ }\hskip 0.2cm $9.0\times 10^{-5}$\mbox{ }\hskip 0.2cm
      &\mbox{ }\hskip 0.2cm $2.6\times 10^{-4}$\mbox{ }\hskip 0.2cm 
      &\mbox{ }\hskip 0.2cm $1.1\times 10^{-4}$\mbox{ }\hskip 0.2cm\\
\hline\hline
\end{tabular}
\end{center}
\end{table}
%

\subsection{Top-quark pair production}
\label{subsec:Top}

An important property of a heavy top\cite{Top,PDG96} is that
it decays before it can form hadronic
bound states\cite{BDKKZ}. This implies in particular that spin effects, 
for instance polarization of, and spin correlations between $t$ and 
$\bar{t}$ quarks  can be analyzed through the distributions and angular
correlations of the weak decay products of the $t$ and $\bar{t}$
quarks. Moreover, these effects can be calculated in perturbation theory.
They provide an additional means for testing SM predictions and 
of searching for possible new physics effects in top quark production
and decay.

The $\gamma t\bar{t}$ vertex  consists of not only the SM tree-level 
vector and axial-vector coupling terms but also a magnetic dipole moment (MDM)
and an EDM coupling. Likewise, in addition to the tree-level SM $Zt\bar{t}$ 
coupling, we have the analogous $Z$ MDM and $Z$ EDM couplings,
of which the latter is called the top-quark weak dipole moment (WDM).  
The MDM-like couplings are present in the SM at the one-loop level.
On the other hand, the EDM-like couplings violate $CP$ and,
due to the structure of the SM, are only present perturbatively
in the SM at the three loop level\cite{Shabalin}. 
In some extensions to the SM such
as left-right models, multi-Higgs-doublet models (MHD), supersymmetric
SM, however, the EDM couplings may be present even at one-loop 
level\cite{Barr,Soni-Xu}. Neglecting the MDM couplings, we assume the 
$\gamma tt$ and $Ztt$ vertices to be given by 
\begin{eqnarray}
\Gamma^V_\mu=v_V\gamma_\mu+a_V\gamma_\mu\gamma_5
            +\frac{c_V}{2m_t}\sigma_{\mu\nu}\gamma_5q^\nu,\qquad
            V=\gamma, Z,
\label{eq:vertex}
\end{eqnarray}
with the vector and axial-vector couplings of the top-quark  given 
in the SM.
Here, $q$ is the four-momentum of the vector boson, $V(=\gamma, Z)$. 
Then, for $m_t=175$ GeV, the top-quark EDM and WDM, $d_{\gamma,Z}$,
are related with $c_{\gamma,Z}$ as  
\begin{eqnarray}
d_{\gamma,Z}=\frac{e}{m_t}c_{\gamma,Z}
            \approx 1.13\times 10^{-16}c_{\gamma,Z}({\rm e}cm).
\end{eqnarray}
%

\subsubsection{Electron-positron mode}
\label{subsubsec:Top_EP}

The angular dependence for the process $e^+e^-\rightarrow t\bar{t}
\rightarrow (X^+b)(X^-\bar{b})$, can be written as
\begin{eqnarray}
\Sigma_{L,R}(\Theta;\theta,\bar{\theta};\phi,\bar{\phi})
  =\sum^{16}_{i=1}{\cal P}_{iL,R}(\Theta)
       {\cal D}_i(\theta,\bar{\theta};\phi,\bar{\phi}).
\end{eqnarray}
where $\Theta$ is the scattering angle for $e^+e^-\rightarrow t\bar{t}$,
and $\theta(\bar{\theta})$ and $\phi(\bar{\phi})$ are the angle
variables for the $b$($\bar{b}$) in the inclusive decays and
for the $l^+$($l^-$) in the semileptonic decays\cite{CJK} of the top and
anti-top quarks, respectively.
The terms, ${\cal P}_\alpha$ and ${\cal D}_\alpha$, can thus be divided 
into four categories under $CP$ and $CP\tilde{T}$\cite{Choi2}.
There exist {\it six} independent $CP$-odd terms among which 
{\it three} terms are $CP\tilde{T}$-even, and the other {\it three}  
$CP\tilde{T}$-odd.

%
\begin{figure}[ht]
\hbox to\textwidth{\hss\epsfig{file=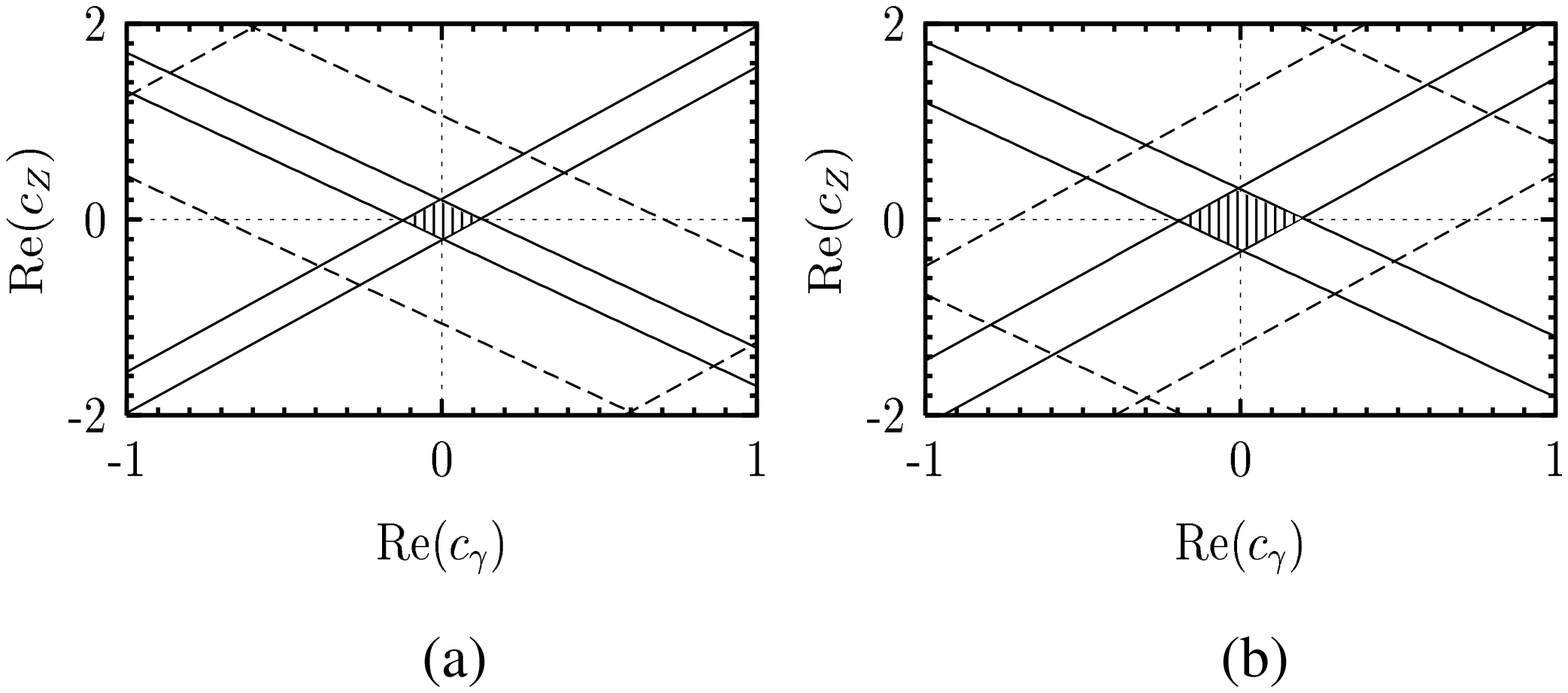,height=3.5cm,width=8cm}\hss}
\vskip 0.3cm
\hbox to\textwidth{\hss\epsfig{file=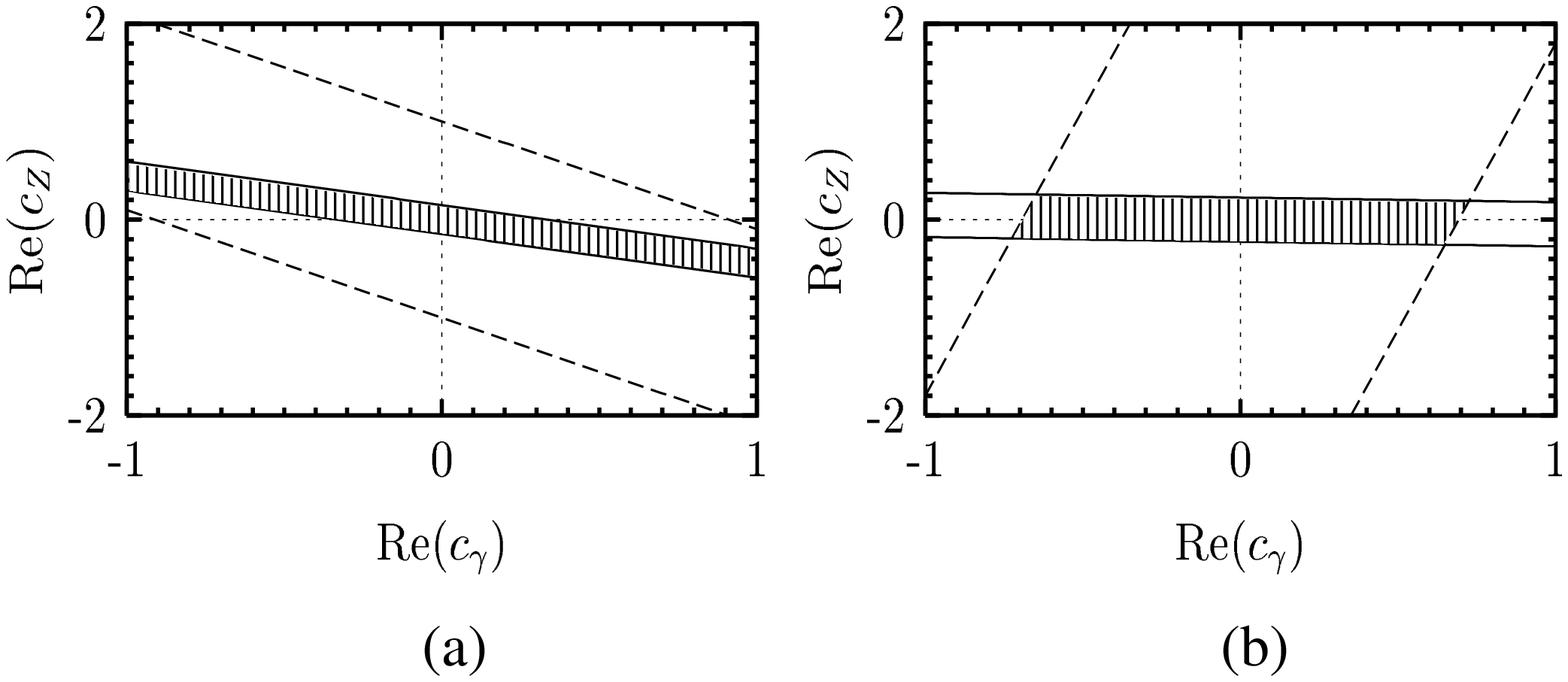,height=3.5cm,width=8cm}\hss}
\caption{The 1-$\sigma$ allowed region of $Re(c_\gamma$) and 
         $Re(c_Z)$ through  (a) $A^b_{1}$ (solid) and 
         $T^b_{33}$ (long-dashed) and (b) $A^l_{1}$ (solid) and 
         $T^l_{33}$ (long-dashed) with the $e^+e^-$ integrated 
         luminosity 10 fb$^{-1}$ for the polarized electron beam (upper part)
         and 20 fb$^{-1}$ for the unpolarized electron beam (lower part), 
         respectively, at $\sqrt{s}=500$ GeV.}
\label{fig:polre}
\end{figure}
%

Including electron beam polarization, Poulose and Rindani\cite{PR}  
recently have considered two new $CP$-odd and $CP\tilde{T}$-even 
asymmetries, of which one asymmetry is essentially equivalent to 
the so-called triple vector product, and two new $CP$-odd and 
$CP\tilde{T}$-odd asymmetries in addition to the two conventional 
lepton energy asymmetries. 
Clearly, that we can use {\it six} more asymmetries among which 
{\it four} asymmetries are $CP$-odd and $CP\tilde{T}$-even and the other 
{\it two} terms are $CP$-odd and $CP\tilde{T}$-odd.

Observables which are constructed from the momenta of the
charged leptons and/or $b$ jets originating from $t$ and $\bar{t}$
decay are directly measurable in future experiments. 
We consider both the inclusive and exclusive semileptonic decays 
\begin{eqnarray}
t\rightarrow bX_{\rm had},\qquad
t\rightarrow bl^+\nu;\ \ l=e,\mu,\tau,
\end{eqnarray}
together with the corresponding charge-conjugated ones, and
we use the $CP\tilde{T}$-even $T_{33}$ and $A_1$ and the $CP\tilde{T}$-odd
$Q_{33}$, $A_2$ and $A_E$ as in the $W$-pair production case. All the
observables are constructed from the momenta
of the final $b$ and $\bar{b}$ in the inclusive decay mode and of the
final $l^+$ and $l^-$ in the exclusive semileptonic decay mode.

Inserting the values of the SM electron vector and axial-vector couplings,
we obtain for left-handed and right-handed electrons
\begin{eqnarray}
c_L=c_\gamma+0.64\delta_Z c_Z,\qquad
c_R=c_\gamma-0.55\delta_Z c_Z,
\label{eq:coupling_num}
\end{eqnarray}
where $\delta_Z=(1-m^2_Z/s)^{-1}$.  
For $m_t=175$ GeV and $m_Z=91.2$ GeV, $1\leq \delta_Z\leq 1.073$.
The $c_Z$ contribution to $c_L$ and $c_R$ is similar in size 
but different in sign.  Naturally, electron polarization plays
a crucial role in discriminating $c_\gamma$ and $c_Z$.

%
\begin{figure}[ht]
\hbox to\textwidth{\hss\epsfig{file=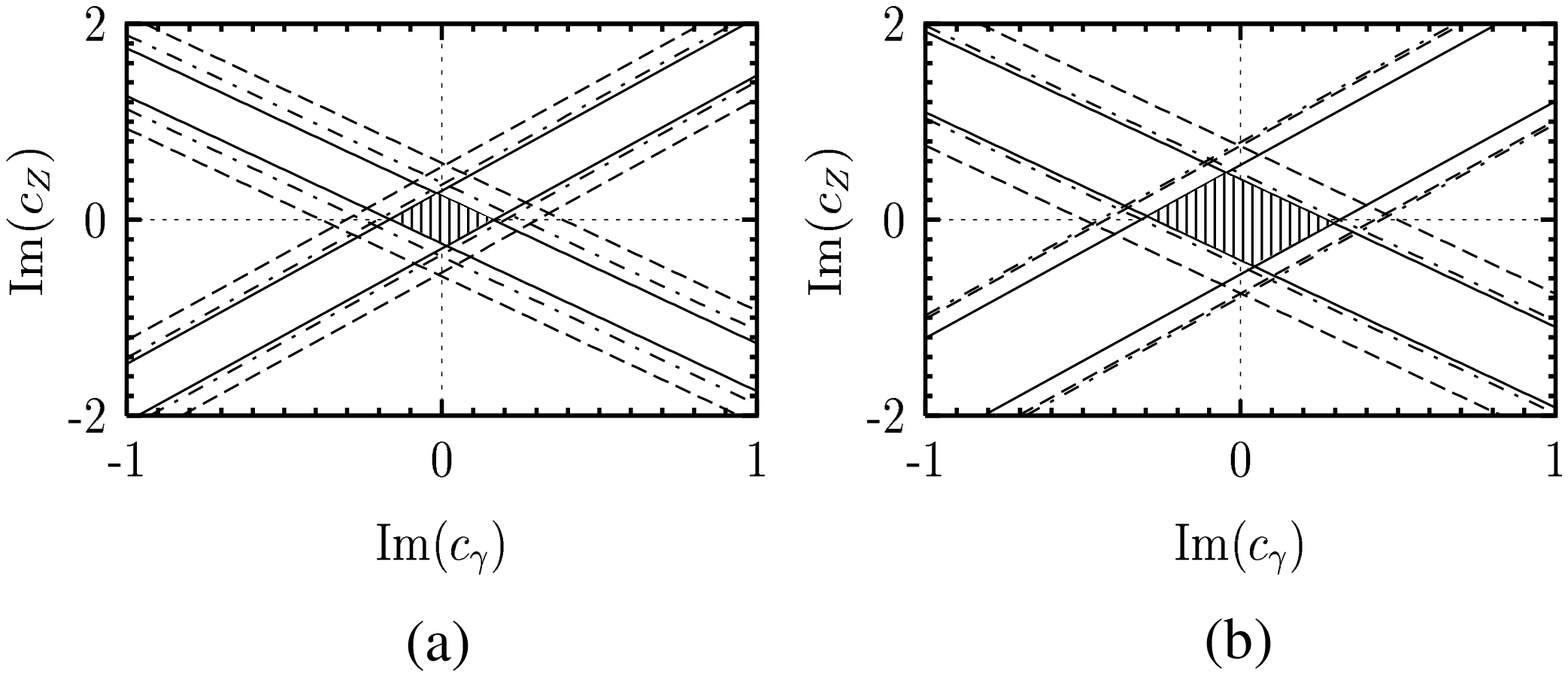,height=3.5cm,width=8cm}\hss}
\vskip 0.3cm
\hbox to\textwidth{\hss\epsfig{file=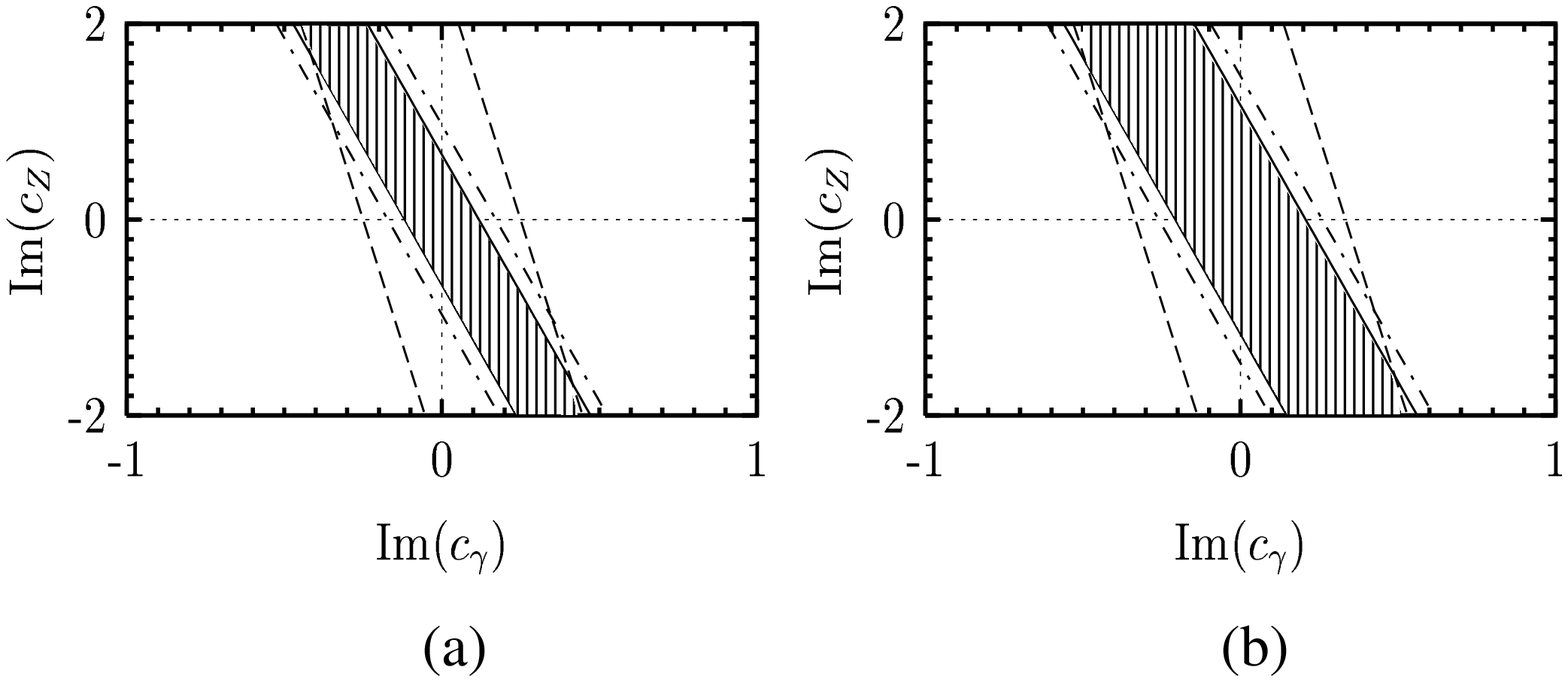,height=3.5cm,width=8cm}\hss}
\caption{The 1-$\sigma$ allowed region of $Im(c_\gamma)$ and 
         $Im(c_Z)$ through (a) $A^b_{E}$ (solid), $A^b_2$ (long-dashed) 
         and $Q^b_{33}$ (dashed) and 
         (b) $A^l_{E}$ (solid), $A^l_2$ (long-dashed) and $Q^l_{33}$ (dashed) 
         with the $e^+e^-$ integrated luminosity 10 fb$^{-1}$ for
         the polarized electron beam (upper part) and 20 fb$^{-1}$ for the 
         unpolarized electron beam (lower part) at $\sqrt{s}=500$ GeV.}
\label{fig:polim}
\end{figure}
%

Our numerical results are presented for the following set of experimental
parameters:
\begin{eqnarray}
\sqrt{s}=0.5\ \ {\rm TeV},\qquad
L_{ee}= \left\{
 \begin{array}{ll}
20 \ {\rm fb}^{-1} & \ \ {\rm for}\ \ {\rm unpolarized}\ \ {\rm electrons} \\
10 \ {\rm fb}^{-1} & \ \ {\rm for}\ \ {\rm polarized}\ \ {\rm electrons} 
\end{array}, \right.
\end{eqnarray}
The shadowed parts in Fig~\ref{fig:polre} show the 1-$\sigma$ allowed 
regions of $Re(c_\gamma$) and $Re(c_Z)$ through (a) $A^b_{1}$ 
and $T^b_{33}$ and (b) $A^l_{1}$ and $T^l_{33}$ with polarized electron 
beams (upper part), respectively, and those in Fig.~\ref{fig:polim} show 
the 1-$\sigma$ allowed regions of $Im(c_\gamma)$ and $Im(c_Z)$ 
through (a) $A^b_{E}$, $A^b_2$ and $Q^b_{33}$ and 
(b) $A^l_{E}$, $A^l_2$ and $Q^l_{33}$ with unpolarized 
electron beams (lower part), respectively. 
We observe several interesting properties from the figures:
(i) The allowed regions strongly depend on electron polarization.    
(ii) Even with unpolarized electrons and positrons, it is possible to obtain
      a closed region for the $CP$-odd parameters by using two or more 
      $CP$-odd asymmetries.  
(iii) With polarized electrons, the most stringent bounds on the
      $CP\tilde{T}$-even and $CP\tilde{T}$-odd parameters are obtained 
      through  $A^b_1$ and $A^b_E$ in the inclusive top-quark decay 
      mode, respectively.
Numerically, the 1-$\sigma$ allowed regions of  
$Re(c_\gamma)$, $Re(c_Z)$, $Im(c_\gamma)$, and $Im(c_Z)$ are 
\begin{eqnarray}
&&|Re(c_\gamma)|\leq 0.12,\qquad |Re(c_Z)|\leq 0.20,\\
&&|Im(c_\gamma)|\leq 0.16,\qquad |Im(c_Z)|\leq 0.27.
\end{eqnarray}
%

\subsubsection{Two-photon mode}
\label{subsubsec:Top_PP}

Extracting ${\cal I}(\Sigma_{00})$ in $\gamma\gamma\rightarrow t\bar{t}$
again does not require identifying the scattering plane, and 
although the $\tau^+\tau^-+\not\!{p}$ modes of 1\% is excluded, 
the remaining 99\% of the events 
can be used to measure ${\cal I}(\Sigma_{00})$.
On the other hand, ${\cal I}(\Sigma_{02})$ turns out to be zero
so that it is useless in determining ${\rm Re}(c_\gamma)$. 

%
 \begin{figure}[ht]
 \hbox to\textwidth{\hss\epsfig{file=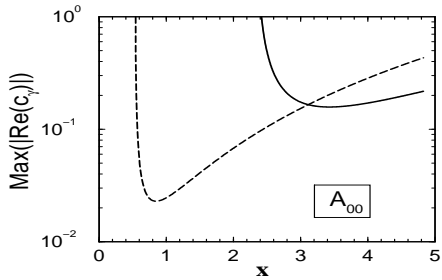,height=4cm,width=6cm}\hss}
 \caption{The $x$ dependence of the $Re(c_\gamma)$ upper bound, 
          Max($|Re(c_\gamma)|$), at $\sqrt{s}=0.5$ (solid) and 
          1 TeV (long-dashed), from $A_{00}$.} 
 \label{fig:rrttsig00}
 \end{figure}
%

We present our numerical results for the same set of experimental 
parameters as the process $\gamma\gamma\rightarrow W^+W^-$.              
Experimentally, $\gamma\gamma\rightarrow W^+W^-$ is the most 
severe background process against $\gamma\gamma\rightarrow t\bar{t}$.
In our analysis, we simply take for a numerical analysis a rather 
conservative value of the detection efficiency $\varepsilon=10\%$, 
even though more experimental 
analysis is required to estimate the efficiency precisely.  
 
It is clear from Fig.~\ref{fig:rrttsig00} and 
Table~\ref{tbl:1-sigma bounds;c_r} that the constraints on 
$Re(c_\gamma)$ through $A_{00}$ are very sensitive to $x$ and 
$\sqrt{s}$.
It is impressive that the doubling of the c.m. energy
enables us to improve the sensitivities almost by a factor of ten.
\begin{table}[ht]
\caption{The optimal 1-$\sigma$ sensitivities to $Re(c_\gamma)$ 
         and their corresponding $x$
         values for $\sqrt{s}=0.5$ and 1 TeV.}
\label{tbl:1-sigma bounds;c_r}
\begin{center}
\begin{tabular}{|c|c|c|}\hline\hline
   $\sqrt{s}$ (TeV)  &   0.5   &  1.0  \\ \hline
       $x$           &   3.43  &  0.85 \\ \hline
   $Re(c_\gamma)$    &   0.16  &  0.02 \\ 
\hline\hline
\end{tabular}
\end{center}
\end{table}
%

\section{$CP$ Violation in the Tau Lepton System}
\label{sec:tau}

The $\tau$ has the same interaction structure as the $e$ and $\mu$ in
the SM, apart from their masses. However for practical purposes\cite{Gomez}
the $\tau$ lepton, the most massive of the known leptons, behaves quite
differently from the $e$ and $\mu$ leptons in that (i)
the $\tau$ has hadronic decay modes (e.g. $\tau\rightarrow \pi\nu,
\rho\nu, a_1\nu, K^*, ...$) which allow an efficient measurement of its
polarization\cite{Tsai} and (ii) the coupling to the neutral
and charged Higgs bosons\cite{Grossman,Falk} and other 
scalar particles is expected to dominate those of the $e$ and $\mu$.
These features allow the $\tau$ to be a rather special experimental 
probe of new physics\cite{Nachtmann1,Kilan,Rindani,Tsai1,Huang,Mirkes}.

\subsection{Tau Lepton EDM}

Recently, the OPAL\cite{OPAL} and ALEPH\cite{ALEPH} detector groups have 
demonstrated by detailed investigations of $Z\rightarrow\tau^+\tau^-$ at 
LEPI that sensitive $CP$ symmetry tests for the $\tau$ WDM at the few per mill 
level can be performed in high energy $e^+e^-$ collisions.
The OPAL group has employed two optimal genuine $CP$-odd observables,
of which one is $CP\tilde{T}$-even and the other one $CP\tilde{T}$-odd,
and the ALEPH group has used the $CP$-odd tensor observable $T_{33}$,
which can give information only on the real part of the $\tau$ WDM.
Specifically the upper limit on the real and imaginary parts
of the $\tau$ WDM obtained at OPAL are 
$|Re(d_Z)|\leq 7.8\times 10^{-18} e{\rm cm}$ and
$|Im(d_Z)|\leq 4.5\times 10^{-17} e{\rm cm}$ with 95\% confidence level,
while the upper limit on the real part of the $\tau$ WDM obtained at ALEPH
is $|Re(d_Z)|\leq 1.5\times 10^{-17} e{\rm cm}$.

Compared with the present constraints on the $\tau$ WDM,
those on the $\tau$ EDM are rather weak and, in contrast to the OPAL
and ALEPH measurements, the reported 
measurements\cite{PDG96} have used indirect methods based on the 
$CP$-even observables such as the differential cross section of 
$e^+e^-\rightarrow \tau^+\tau^-$\cite{Delaguila}, the 
partial decay widths of $Z\rightarrow \tau^+\tau^-$\cite{Grifols} and
$Z\rightarrow \tau^+\tau^-\gamma$\cite{Escribano}. 
We claim that {\it the measurements may not be regarded as genuine
$\tau$ EDM measurements and therefore the quoted values have to be 
replaced by those by  direct measurements through $CP$-odd 
observables}\cite{Nachtmann}. 
Surprisingly, no direct $\tau$ EDM measurements have been reported
in the literature.
Therefore, we strongly suggest the $\tau$ EDM to be measured directly
at the existing collider facilities TRISTAN, LEPII and CLEOII.

In the present report, first of all, we present a rough comparison of 
the potential of TRISTAN and LEPII in the $\tau$ EDM
measurements based on the following experimental parameters:
\begin{eqnarray}
\begin{array} {clll}
{\rm TRISTAN} & :  &  \sqrt{s}=\ \ 60 {\rm GeV},   
              & \ \ {\cal L}_{ee}=271 {\rm pb}^{-1},\\
{\rm LEPII}   & :  &  \sqrt{s}=180 {\rm GeV},
              & \ \ {\cal L}_{ee}=500 {\rm pb}^{-1}
\end{array}
\label{tau:exp_par}
\end{eqnarray}
In the above quoted values, we note that TRISTAN
has already accumulated the integrated luminosity 
$271$ pb$^{-}$\cite{Hanai}, but LEPII is assumed to eventually 
accumulate 500 pb$^{-1}$, which is the designed integrated
luminosity per year. 

\begin{figure}[ht]
\hbox to\textwidth{\hss\epsfig{file=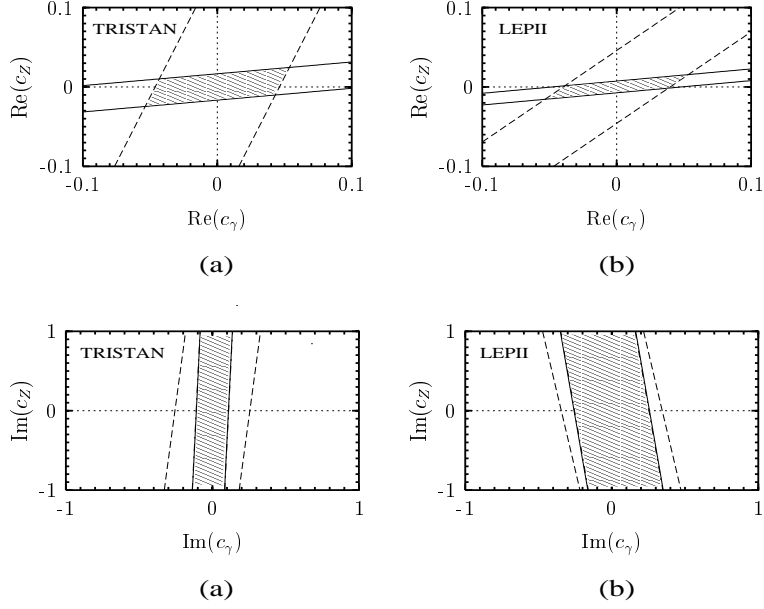,
                   height=8cm,width=10cm}\hss}
\vskip 0.5cm
\caption{The 1-$\sigma$ allowed region (left part) of 
         $Re(c_\gamma)$ and $Re(c_Z)$
         through $A_1$ (solid) and $T_{33}$ (long-dashed), 
         and that (right part) of $Im(c_\gamma)$ and $Im(c_Z)$ through
         $A_E$ (solid) and $A_2$ (long-dashed) at TRISTAN and 
         LEPII with the experimental parameter values 
         (\ref{tau:exp_par}). The lines for $Q_{33}$ are overlapped
         with those for $A_E$.}
\label{fig:Re_EDM}
\end{figure}

For the sake of simplicity, we consider the simplest semileptonic decay
of the $\tau$ lepton, $\tau\rightarrow\pi\nu_\tau$ whose branching 
ratio is 11\%. Certainly, to obtain better constraints, all the 
reconstructable decay channels of the $\tau$ should be included.
Given the branching fraction (11\%), TRISTAN and LEPII yield
ninety-seven and twenty-three events of the sequential
process $e^+e^-\rightarrow \tau^+\tau^-\rightarrow
(\pi^+\bar{\nu}_\tau) (\pi^-\nu_\tau)$.
To make a rough estimate of the sensitivities to be obtained, we
use the same set of $CP$-odd observables as used in the top-quark
pair production at NLC in Section~\ref{sec:NLC}, that is to say, 
the $CP\tilde{T}$-even $A_1$ 
and $T_{33}$ and the $CP\tilde{T}$-odd $A_E$, $A_2$, and $Q_{33}$.
All the $CP$-odd observables are constructed from the momenta of the
electron in the initial state and two pions in the final state.

We exhibit in Fig.~\ref{fig:Re_EDM} the 1-$\sigma$ allowed 
region of the $CP$-violation parameters $c_\gamma$ and $c_Z$, which are 
numerically related with the $\tau$ EDM and WDM as follows
\begin{eqnarray}
d^{\gamma,Z}_\tau=1.1\times 10^{-14} c_{\gamma, Z} (e{\rm cm}).
\end{eqnarray}
Incidentally, it is clear that TRIATAN and LEPII can not compete with LEPI
in measuring the $\tau$ WDM as can be seen clearly by comparing the
bound in the figure and the values quoted in the first paragraph of the
present section. So, we take into account the LEPI measurements
and then we derive the 1-$\sigma$ allowed range for the 
$\tau$ EDM by taking the $x$-axis cut values. Before listing the 
numerical values, we note that the $CP\tilde{T}$-odd 
observables $A_E$ and $Q_{33}$ provide very similar constraints
on the imaginary part of the $\tau$ EDM so that their 1-$\sigma$
boundary lines can be hardly distinguished in Fig.~\ref{fig:Re_EDM}. 
Combined with the LEPI results, 
the 1-$\sigma$ sensititivies to the $\tau$ EDM are
\begin{eqnarray}
\begin{array} {clll}
{\rm TRISTAN} & : &  |Re(d_\tau^\gamma)|\leq 5.3\times 10^{-16} (e{\rm cm}),
                  &  |Im(d_\tau^\gamma)|\leq 1.1\times 10^{-15} (e{\rm cm}),\\
{\rm LEPII}   & : &  |Re(d_\tau^\gamma)|\leq 4.4\times 10^{-16} (e{\rm cm}),
                  &  |Im(d_\tau^\gamma)|\leq 2.8\times 10^{-15} (e{\rm cm}),
\end{array}
\end{eqnarray}
TRISTAN and LEPII can measure the real part of the $\tau$ EDM with 
similar sensitivities, which may become comparable with those quoted by 
the PDG group, in the case that all the reconstructable decay modes of the 
tau lepton are included.  

\begin{figure}[ht]
\hbox to\textwidth{\hss\epsfig{file=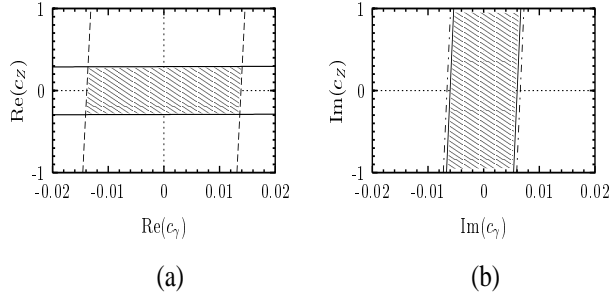,height=4cm,width=8cm}\hss}
\vskip 0.5cm
\caption{(a) The 1-$\sigma$ allowed region of 
         $Re(c_\gamma)$ and $Re(c_Z)$ through $A_1$ (solid) and 
         $T_{33}$ (long-dashed) and (b) that  of 
         $Im(c_\gamma)$ and $Im(c_Z)$ through $A_E$ (solid)
         and $Q_{33}$ (dot-dashed) with the $3.02\times 10^{6}$ 
         $\tau$ pairs at the CLEOII energy $\sqrt{s}=10.6$ GeV.}
\label{fig:cleo}
\end{figure}

On the other hand, the $e^+e^-$ storage ring CESR has accumulated the 
integrated luminosity of about $3.5$ fb$^{-1}$ at the c.m. energy 
$\sqrt{s}=10.6$ GeV in the years between 1990 and 1994\cite{CLEOII}. 
It corresponds to the production of about $3.02\times 10^{6}$ $\tau$ pairs, 
which is about four-hundred times larger than the number of $\tau$ pairs 
at TRISTAN with its integrated luminosity 271 pb$^{-1}$. 
Therefore, a great improvement of the sensitivities to the $\tau$ EDM 
is expected. As a matter of fact, as shown in Fig.~\ref{fig:cleo}, 
the 1-$\sigma$ allowed range for $Re(c_\gamma)$ is about four times smaller 
than that at TRISTAN and LEPII, while the 1-$\sigma$ allowed range of
$Im(c_\gamma)$ is about fifteen times smaller than that at TRISTAN.
Quantitatively, the 1-$\sigma$ sensitivities to the $\tau$ EDM, if
the LEPI results on the $\tau$ WDM are included, are
\begin{eqnarray}
|Re(d^\gamma_\tau)|\leq 1.5\times 10^{-16} (e{\rm cm}),\qquad
|Im(d^\gamma_\tau)|\leq 6.7\times 10^{-17} (e{\rm cm}).
\end{eqnarray}
%

%
%
 
The quoted indirect limits of the absolute value of the $\tau$ EDM is
$5\times 10^{-17} (e {\rm cm})$, which is still three times smaller
than the expected limits from the CLEOII measurements.
However, we should note that only the simplest decay, 
$\tau\rightarrow\pi\nu_\tau$, whose branching fraction is 11\%, 
has been considered in our analysis. Therefore, we expect that including 
other dominant decay modes such as $\tau\rightarrow \rho\nu_\tau, 
a_1\nu_\tau,\ldots$ enable us to easily improve the sensitivities 
more than five times. The more detailed work on the CLEOII measurements of
the $\tau$ EDM with all the reconstructable decay modes included
is in progress\cite{Choi4}.

\subsection{$CP$ Violation in Semileptonic $\tau$ Decays}

An observation of $CP$ violation in $\tau$ decays requires the existence 
of not only a $CP$ violating phase but also the interference of processes with 
different $CP$ phases. One can have in general a $CP$ violating phase 
between the $W$-exchange diagram and the charged-scalar-exchange 
diagram in models beyond the SM such as MHD and scalar
leptoquark (SLQ) models. On the other hand, two resonance states 
with large width-to-mass ratios\cite{PDG96}, which decay to the same 
final states, leads to an interference of the two processes with different 
$CP$ phases. 

The decay amplitudes of $\tau\rightarrow 3\pi\nu_\tau$ and
$\tau\rightarrow K\pi\nu_\tau$ have contributions from the two 
overlapping resonances ($a_1,\pi^\prime$) and ($K^*,K^*_0$)
with different spins and relatively large width-to-mass 
ratios. Here we should note that the parameters of the resonances
are not so accurately determined. In the $\tau$ decays, various 
phenomenological parameterizations\cite{Jadach,Tornqvist,Isgur,Decker}
of the form factors have been 
employed to analyze experimental data. Keeping in mind the
uncertainty of the resonance parameters, we simply adopt the
parameterization of the $\tau$-decay library TAUOLA\cite{Jadach} 
for the masses and widths of the resonances tabulated as follows
\begin{center}
\begin{tabular}{|c||c|c||c|c|}\hline\hline
resonances  &   $a_1$     &    $\pi^\prime$   &  $K^* $   &   $K^*_0$ \\ \hline
$J^P$       &   $1^+$     &    $0^-$          &  $1^-$    &   $0^+$   \\ \hline
mass (GeV)  &   $1.25$    &    $1.30$         &  $0.89$   &   $1.43$  \\ \hline
width (GeV) &   $0.6$     &    $0.3$          &  $0.05$   &   $0.29$  \\ 
 \hline\hline
\end{tabular}
\end{center}

The semileptonic decay modes of the $\tau$ are promising for 
the detection of $CP$ violation for the following reasons. 
First, no tagging of the other $\tau$ is necessary. 
Second, the decay modes can be measured not only at the CESR 
and CERN $e^+e^-$ Collider LEP but
also at the planned B factories and $\tau$-charm factories\cite{Gomez} 
where many $\tau$ leptons ($10^7$ to $10^8$) are expected to be produced.

There is, however, an experimental difficulty; Since the neutrinos 
escape detection, it is in general very difficult to reconstruct 
the $\tau$ rest frame. Nevertheless, there are two situations
where the $\tau$ rest frame can be actually reconstructed. 
One is $\tau$-pair production close to threshold where $\tau$ leptons
are produced at rest. This possibility can be realized at future
$\tau$-charm factories. The other is when both $\tau$ leptons decay 
semileptonically. In the latter case impact parameter methods allow us
to reconstruct the $\tau$ rest frame even for $\tau$'s in flight.
Obviously, the impact parameter method\cite{Kuhn} 
requires the full identification
of the decay product momenta. Therefore, we use the decay modes
$\tau^-\rightarrow \pi^-\pi^+\pi^-\nu_\tau$ in the $3\pi$ mode and 
$\tau^-\rightarrow (K^0_s\rightarrow\pi^+\pi^-)\pi^-\nu_\tau$ in
the $K\pi$ mode, of which
the branching fractions are 6.8\% and 0.33\%, respectively.
 
Generally, the matrix element for the semileptonic $\tau$ decays 
can be cast into the form
\begin{eqnarray}
M=\sqrt{2}G_F \Bigl[(1+\chi)\bar{u}(k,-)\gamma^\mu P_- u(p,\sigma)J_\mu
                 +\eta\bar{u}(k,-)P_+ u(p,\sigma)J_S\Bigr],
\label{decaym}
\end{eqnarray}
where $G_F$ is the Fermi constant, $p$ and $k$ are the four momenta 
of the $\tau$ lepton and the tau neutrino, respectively, and $\eta$ is 
a parameter determining the size of the scalar contribution.
$J_\mu$ and $J_S$ are the vector hadronic current and 
the scalar hadronic current given by
\begin{eqnarray}
\begin{array}{ll}
J^{3\pi}_\mu
     =\cos\theta_C\langle 3\pi|\bar{d}\gamma_\mu(1-\gamma_5)u|0\rangle, &
\ \ J^{3\pi}_S  
     =\cos\theta_C\langle 3\pi|\bar{d}(1+\gamma_5)u|0\rangle, \\
J^{K\pi}_\mu
     =\sin\theta_C\langle (K\pi)^-|\bar{s}\gamma_\mu u|0\rangle, & 
\ \ J^{K\pi}_S
     =\sin\theta_C\langle (K\pi)^-|\bar{s}u|0\rangle,
\end{array}
\end{eqnarray}
where $\theta_C$  is the Cabibbo angle.
The explicit form of $J_\mu$  is found in the $\tau$ decay library TAUOLA
and the hadronic scalar current $J_S$ can be determined by the
Dirac equation from the $J_\mu$ current.

We can now construct a $CP$-even sum $\Sigma$
and a $CP$-odd difference $\Delta$ of the differential $\tau^\pm$ 
decay rates:
\begin{eqnarray}
&& \Sigma=\frac{{\rm d}\Gamma}{{\rm d}\Phi_{X^-}}
                [\tau^-\rightarrow X^-\nu_\tau]
         +\frac{{\rm d}\Gamma}{{\rm d}\Phi_{X^+}}
                [\tau^+\rightarrow X^+\bar{\nu}_\tau],
          \nonumber\\ 
&& \Delta=\frac{{\rm d}\Gamma}{{\rm d}\Phi_{X^-}}
                [\tau^-\rightarrow X^-\nu_\tau]
         -\frac{{\rm d}\Gamma}{{\rm d}\Phi_{X^+}}
                [\tau^+\rightarrow X^+\bar{\nu}_\tau]\Bigr\},
\end{eqnarray}
where $X=3\pi, K\pi$ and $\Phi_{X^\pm}$ are the full kinematic variables
of the detectable final states $X^+$ and $X^-$, which should be
related with each other under $CP$ transformations.
Then, the $CP$-odd $\Delta$ is proportional to the imaginary
part of the parameter $\xi$, which is given by
\begin{eqnarray}
\xi_{3\pi}=\frac{m^2_{\pi^\prime}}{(m_u+m_d)m_\tau}
    \left(\frac{\eta}{1+\chi}\right),\qquad
\xi_{K\pi}=\frac{m^2_{K^*_0}}{(m_s-m_d)m_\tau}
    \left(\frac{\eta}{1+\chi}\right).
\end{eqnarray}
We re-emphasize that every $CP$ asymmetry requires not only 
a non-vanishing $Im(\xi)$ but also the interference between 
the longitudinal mode of the vector
meson and the scalar meson and the interference is proportional to the 
decay constants $f_{\pi^\prime}$ and $f_{K^*_0}$. The value of the 
$\pi^\prime$ decay constant, $f_{\pi^\prime}=0.02$-$0.08$ GeV 
estimated in Ref.~\cite{Isgur} and $f_{\pi^\prime}=0.02$ GeV quoted 
in TAUOLA are quitely likely
invalid because the mixing between the chiral pion field and a
massive pseudoscalar $q\bar{q}$ bound state should be considered. 
So, we have reconsidered $f_{\pi^\prime}$ in the chiral Lagrangian
framework and have shown that it actually vanishes in the chiral
limit due to the mixing of $\pi^\prime$ with the chiral pion 
field\cite{Choi3}.
$f_{\pi^\prime}$ is indeed proportional to the square of the pion mass
$m_\pi$ and so it may be much smaller than the value quoted in 
TAUOLA. This small $f_{\pi^\prime}$ value is also supported by
arguments from QCD sum rules. 
On the other hand, $f_{K^*_0}$ can be well determined 
phenomenologically by using the well-measured $K^*_0$ width 
and the QCD sum rule technique.
Considering possible uncertainties in our estimates we use in our
numerical analysis $f_{\pi^\prime}=(1\sim 5)\times 10^{-3}$ GeV
for the $\pi^\prime$ decay constant, which is a slightly broader range
than estimated, and $f_{K^*_0}=45$ MeV
for the $K^*_0$ decay constant, which is rather well estimated. 

As possible new sources of $CP$ violation detectable in the $\tau$ 
decays we consider new scalar-fermion interactions which preserve 
the symmetries of the SM. Under those conditions there exist
only four types of scalar-exchange models\cite{Davies}, 
which can contribute to the decays, $\tau\rightarrow 3\pi\nu_\tau$ and
$\tau\rightarrow (K\pi)\nu_\tau$.
One is the MHD model\cite{Grossman} and the other three are 
scalar-leptoquark (SLQ) models\cite{Davies,Hall}. 
The quantum numbers of the three leptoquarks under the gauge group 
${\rm SU}(3)_C\times {\rm SU}(2)_L\times {\rm U}(1)_Y$ are 
\begin{eqnarray}
&& \Phi_1=\left(3,3,\ \ \frac{7}{6}\right)\ \ ({\rm model}\ \ 
         {\rm I}),\nonumber\\
&& \Phi_2=\left(3,1,-\frac{1}{3}\right)\ \ ({\rm model}\ \ 
         {\rm II}),\nonumber \\
&& \Phi_3=\left(3,3,-\frac{1}{3}\right)\ \ ({\rm model}\ \ 
         {\rm III}),
\end{eqnarray}
respectively. The hypercharge $Y$ is defined to be $Q=I_3+Y$.

The constraints on the $CP$-violation parameters $\xi_{3\pi}$ and $\xi_{K\pi}$
depend upon the values of the $u$, $d$ and $s$ current quark masses,
which are taken to be
\begin{eqnarray}
m_u=5\ \ {\rm MeV},\qquad m_d=9\ \ {\rm MeV},\qquad  
m_s=320\ \ {\rm MeV}.
\end{eqnarray}
The present experimental constraints on the $CP$-violation parameters in the
MHD model, which have been extensively reviewed in Ref.\ \cite{Grossman},
are 
\begin{eqnarray}
  \begin{array}{l}
\Gamma(B\rightarrow X\tau\nu_\tau)\\
\Gamma(K^+\rightarrow\pi^+\nu\bar{\nu})
  \end{array}
\Rightarrow 
|Im(\xi_{MHD})|_{3\pi}<0.28, \ \
|Im(\xi_{MHD})|_{K\pi}<0.48,
\label{HDc}
\end{eqnarray}
for the charged Higgs mass $M_H=45$ GeV.
There are at present no direct constraints on the $CP$-violating 
parameters in the SLQ models. Assuming that the three leptoquark couplings 
to fermions are universal,  
we can roughly estimate the constraints on the SLQ $CP$-violating 
parameters. The constraints\cite{PDG96,Dohmen,Marciano} 
on the $CP$-violation parameters 
from various low energy experiments then are
\begin{eqnarray}
\Gamma(K_L\rightarrow \mu e)
   &\Rightarrow& 
   |Im(\xi^I_{SLQ})|_{3\pi} < 1.5\times 10^{-3}, \nonumber\\
\frac{\Gamma(\mu{\rm Ti}\rightarrow e{\rm Ti})}{\Gamma
      (\mu{\rm Ti}\rightarrow {\rm capture})}
   &\Rightarrow& 
   |Im(\xi^{II}_{SLQ})|_{3\pi} < 0.9\times 10^{-3}, \\
D\bar{D}\ \ {\rm mixing} 
   &\Rightarrow& \left\{\begin{array}{l} 
    |Im(\xi^I_{SLQ})|_{K\pi} < 4\times 10^{-2}, \\ 
    |Im(\xi^{II}_{SLQ})|_{K\pi}< 6\times 10^{-2},
        \end{array}\right. \nonumber\\
\Gamma(K\rightarrow \pi\nu\bar{\nu})
   &\Rightarrow&
    |Im(\xi^{III}_{SLQ})|_{K\pi}< 10^{-5},
\end{eqnarray}
Compared to the constraint on the $Im(\xi_{MHD})$,
the constraints on the SLQ $CP$-violation parameters
are much more severe. Especially, the constraint on $Im(\xi^{III}_{SLQ})$
is so severe that no $CP$-violation effects are expected to be observed.
Therefore we will not consider this type-III SLQ model any longer.

The following numerical analysis is made for the maximally allowed values 
of the $CP$-violation parameters and the number of $\tau$ leptons required 
to detect $CP$ violation is estimated by using 
the optimal $CP$-odd asymmetry\cite{Atwood}  
\begin{eqnarray}
w_{\rm opt}=\frac{\Delta}{\Sigma}.
\end{eqnarray}
\vskip 0.5cm
\begin{table}[ht]
\caption{ The maximal expected size of $\varepsilon_{\rm opt}$ and 
the number of $\tau$ leptons, $N$, for detection with 
the $\varepsilon_{\rm opt}$ at the $2\sigma$ level in the $3\pi$
and $K\pi$ modes for $f_{\pi^\prime}=1\sim 5$ MeV and $f_{K^*_0}=45$ MeV 
with the maximally-allowed values for the $CP$-violation parameters.} 
\begin{center}
\begin{tabular}{|c|c|c|c|c|}\hline\hline
Model  & $\varepsilon^{3\pi}_{\rm opt}$(\%)
       & $\varepsilon^{K\pi}_{\rm opt}$(\%) 
       & $N^{3\pi}$                     & $N^{K\pi}$  \\ \hline
MHD    & $0.13\sim 0.67$                & $2.3$  
       & $(0.13\sim 3.3)\times 10^7$    & $2.3\times 10^6$  \\ \hline
SLQI   & $(0.7\sim 3.6)\times 10^{-3}$  & $0.2$  
       & $(0.5\sim 11)\times 10^{11}$   & $3.4\times 10^8$ \\ \hline
SLQII  & $(0.4\sim 2.2)\times 10^{-3}$  & $0.3$  
       & $(1.2\sim 33)\times 10^{11}$   & $1.5\times 10^8$ \\ 
\hline\hline
\end{tabular}
\end{center}
\label{tab:size}
\end{table}

Table~\ref{tab:size} shows the expected size of
$\varepsilon_{\rm opt}$, along with the number of $\tau$  
leptons, $N$, required to obtain the 2-$\sigma$ signal with the
optimal asymmetry, $\varepsilon_{\rm opt}$, 
in the MHD model and the two SLQ models for the $CP$-violation 
parameter values. The values of $\varepsilon_{\rm opt}$ and 
the corresponding $N$ values in Table~\ref{tab:size} show that 
the $CP$-violating effects from the MHD model can be detected with 
less than $10^7$ $\tau$ leptons, while still more than $10^{8}$ $\tau$ 
leptons are required to see $CP$ violation in the SLQ models. 
In light of the fact that about $10^7$ and $10^8$ $\tau$ leptons are 
produced yearly at $B$ and $\tau$-Charm factories, respectively, 
$CP$ violation from the MHD model can hopefully be observed.

\section{Summary and Conclusion}
\label{sec:Conclusion}

Let us summarize the results presented in the present report.

First of all, we have made a systematic study of observable experimental
$CP$-odd distributions in the $e^+e^-$ annihilation processes,
$e^+e^-\rightarrow W^+W^-$ and $e^+e^-\rightarrow t\bar{t}$ and of initial
$CP$-odd two-photon polarization configurations in the two-photon
fusion processes, $\gamma\gamma\rightarrow W^+W^-$ and
$\gamma\gamma\rightarrow t\bar{t}$ at NLC, which could serve as 
tests of possible anomalous $CP$-odd three-boson and four-boson 
couplings and top-quark EDM and WDM couplings.

We have found that backscattering linearly polarized Compton
laser light off the electron or positron beam allows us to prepare 
two types of initial $CP$-odd two-photon polarization 
configurations and that the sensitivities through the two $CP$-odd
asymmetries to the $CP$-violation parameters depend strongly on the 
parameter $x$, i.e. the laser beam frequency.
So, from an experimental point of view, it is very crucial to be able to 
adjust the initial laser beam frequency in probing $CP$ violation.

Several works have shown that new models beyond the SM
such as MHD and MSSM could have $CP$-odd parameters of the
size\cite{WEDM} which are close to the limits obtainable from $W$-pair 
production both in $e^+e^-$ annihilation and two-photon fusion at an NLC.
On the other hand, the expected size of the top-quark EDM and WDM
is smaller than the reachable limits by an order of magnitude.
However, it is expected that the two-photon mode with higher c.m. energies 
and adjustable laser frequencies could approach very close to the 
expectation.

Secondly, we have pointed out that no direct measurements of the $\tau$
EDM have been reported, while two very good measurements of the $\tau$ 
WDM have been performed by OPAL\cite{OPAL} and ALEPH\cite{ALEPH}.
For that reason, we have suggested any direct measurements of the $\tau$
EDM to be immediately presented at the presently existing collider 
facilities, TRISTAN, LEPII and CLEOII.
We have compared the potential of TRISTAN, LEPII and CLEOII in the 
$\tau$ EDM measurements by using the simplest $\tau$ decay mode 
$\tau\rightarrow \pi\nu_\tau$.
As a result, we have found that, when all the constructible $\tau$ 
decay modes are included, CLEOII should enable us to obtain limits
similar to or better than the indirect limits quoted in the 
literature\cite{PDG96}. 

Thirdly, the semileptonic $\tau$ decays,
$\tau\rightarrow 3\pi\nu_\tau$ and $\tau\rightarrow K\pi\nu_\tau$,
involve two different intermediate resonances with large ratios
of widths to masses so that we have proposed to investigate them
as very promising decay channels to probe $CP$ violation which stems
from new sources completely different from the SM $CP$-violation mechanism.
Quantitatively, it turns out that the $CP$-violating effects due to the
charged Higgs exchanges in the MHD model might be detected at the planned
$B$ and proposed $\tau$-charm factories, while those in the SLQ models
are hardly detectable.

Consequently, we conclude that (i) $CP$ violation from new physics 
can constitute an interesting research programme at NLC,
(ii) any direct $\tau$ EDM measurements should be made immediately
at the presently existing collider experiments, in particular at CLEOII
to replace the indirect limits quoted in the literature, and
(iii) the semileptonic $\tau$ decays are worthwhile to be investigated
as promising channels to probe $CP$ violation from new physics.
Any $CP$-violation phenomenon in the processes under consideration,
if discovered, implies new $CP$-violation mechanism, which can have
far-reaching physical consequences.

\section*{Acknowledgments}

The authors would like to thank Profs.~O.~Nachtmann and
M.~Tanabashi for useful comments and fruitful discussions.
SYC thank the KEK Theory Group and the DESY theory group for
their kind hospitality during his stay.
The work of SYC was supported in part by the KOSEF-DFG large collaboration 
project, Project No. 96-0702-01-01-2 and in part by the Korean Science 
and Engineering Foundation (KOSEF) and Korean Federation of Science and 
Technology Societies through the Brain Pool program
and the work of MSB was supported in part by KOSEF and Research University
Fund of College of Science at Yonsei university supported by 
the Ministry of Education, Korea.

\newcommand{\prd}[1]{Phys.~Rev.~D{{\bf #1}}}
\newcommand{\prl}[1]{Phys.~Rev.~Lett.~{{\bf #1}}}
\newcommand{\plb}[1]{Phys.~Lett.~{{\bf #1B}}}
\newcommand{\npb}[1]{Nucl.~Phys.~{{\bf B#1}}}
\newcommand{\zpc}[1]{Z.~Phys.~{{\bf C#1}}}
\newcommand{\progtp}[1]{Prog.~Theor.~Phys.~{{\bf #1}}}
\newcommand{\jetpl}[1]{JETP Lett.~{{\bf #1}}}
\newcommand{\sjnp}[1]{Sov.~J.~Nucl.~Phys.~{{\bf #1}}}


\begin{thebibliography}{99}
%
\bibitem{Glashow} S.L.~Glashow, Nucl. Phys. {\bf 22}
    (1961) 579; S.~Weinberg, \prl{19} (1967) 1264;
   A.~Salam, in {\it Elementary Particle Theory}, ed.~N.~Svartholm 
   (Almquist and Wiksells, Stockholm, 1969), p.~367.
%
\bibitem{EFT} H.~Georgi, Ann.~Rev.~Nucl.~Part.~{\bf 43} (1994) 209;
    \npb{361} (1991) 339.
%
\bibitem{Christenson} J.~H.~Christenson, J.~W.~Cronin, 
   V.~L.~Fitch, and R.~Turlay, \prl{13} (1964) 138.
%
\bibitem{Cabibbo} N.~Cabibbo, \prl{10} (1963) 531.
%
\bibitem{Kobayashi} M.~Kobayashi and T.~Maskawa, \progtp{49}
   (1973) 652.
%
\bibitem{Wolfenstein} L.~Wolfenstein, 
   \prl{13} (1964) 562.
%
\bibitem{Kolb} E.~D.~Kolb and M.~S.~Turner, 
  {\it The Early Universe} (Addison-Wesley-Publishing, New York, 1990).
%
\bibitem{Sakharov} A.~D.~Sakharov, ZhETF Pis.~Red.~5 (1967) 32;
   \jetpl{5} (1967) 24.
%
\bibitem{Huet} G.R.~Farrar and M.E.~Shaposhnikov, 
   \prd{50} (1994) 774; M.B.~Gavela {\it et al.}, \npb{430} (1994) 382;
   P.~Huet and E.~Sather, \prd{51} (1995) 379.
%
\bibitem{Shabalin} E.~P.~Shabalin, \sjnp{28} (1978) 75;
   J.~Donoghue, \prd{18}, (1978) 1632; I.B.~Khriplovich, \sjnp{44}
   (1986) 659; \plb{173} (1986) 193; A.~Czarnecki and B.~Krause, 
   Acta.~Phys.~Pol.~B {\bf 28} (1997) 829; \prl{78} (1997) 4339.               %
\bibitem{Hoogeveen} F.~Hoogeveen, \npb{341} (1990) 322.
%
\bibitem{Choi1} S.Y.~Choi, K.~Hagiwara, and M.S.~Baek, \prd{54} 
   (1996) 6703; M.S.~Baek and S.Y.~Choi, Preprint No. YUMS 97-5,
   SNUTP 97-028.
%
\bibitem{Choi2} S.Y.~Choi and K.~Hagiwara, \plb{359} (1995) 369;
   M.S.~Baek, S.Y.~Choi and C.S.~Kim, Preprint No. YUMS 97-7, 
   SNUTP 97-035, hep-ph/9704312.
%
\bibitem{Choi3} S.Y.~Choi, K.~Hagiwara and M.~Tanabashi,
   \prd{52} (1995) 1614.
%
\bibitem{NLC} Proceedings of the 1st International Workshop
   on ``Physics and Experiments with Linear $e^+e^-$ Colliders"
   (Saariselka, Filand, September 1991),
   eds. R.~Orava, P.~Eerola, and M.~Nordberg
   (World Scientific, Singapore, 1992);
   Proceedings of the 2nd International Workshop
   on ``Physics and Experiments with Linear $e^+e^-$ Colliders"
   (Waikoloa, Hawaii, April 1993),
   eds. F.A.~Harris, S.L.~Olsen, S.~Pakvasa, and X.~Tata 
   (World Scientific, Singapore, 1993),
   Proceedings of the 3rd International Workshop
   on ``Physics and Experiments with Linear $e^+e^-$ Colliders"
   (Iwate, Japan, September 1995),
   eds. A.~Miyamoto, Y.~Fujii, T.~Matsui, and S.~Iwata,
   (World Scientific, Singpore, 1996).
%
\bibitem{GKS} I.F.~Ginzburg, G.L.~Kotkin, V.G.~Serbo, S.L.~Panfil
   and V.I.~Telnov, Sov.~ZhETF Pis'ma {\bf 34}, 514 (1981) [JETP 
   Lett.~{\bf 34}, 491 (1982)]; Nucl.~Instr.~and~Meth. {\bf 205}, 47
   (1983); I.F.~Ginzburg, G.L.~Kotkin, S.L.~Panfil, 
   V.G.~Serbo, and V.I.~Telnov, {\it ibid} {\bf 219}, 5 (1984);
   V.I.~Telnov, {\it ibid} {\bf A294}. 72 (1992); 
   J.F.~Gunion and H.E.~Haber, Phys.~Rev.~D {\bf 48} (1993) 5109;
   S.Y.~Choi and F.~Schrempp, \plb{272}, 149 (1991);
   E.~Yehudai, \prd{44}, 3434 (1991); Ph.~D Thesis,
   SLAC-PUB-383.
%
\bibitem{Gomez} J.J.~Gomez-Cadenas, in {\it Proceedings of 
   the Third Workshop on the Tau-Charm Factory}, Marbella, Spain, 1993,
   edited by J.~Kirkby and R.~Kirkby (Editions Frontiers, Gif-sur-Yvette, 
   France, 1994); B.C.~Barish and R.~Stroynowski, Phys.~Rep.~{\bf 157} 
   (1988) 1; K.~Riles, Int.~J.~Mod.~Phys.~{\bf A7} (1992) 7647;
   M.L.~Perl, Rep.~Prog.~Phys.~{\bf 55} (1992) 653.
%
\bibitem{WbAb} 
   W.~Bernreuther and A.~Brandenburg, \plb{314} (1993) 104; 
   W.~Bernreuther, J.~P.~Ma, and B.~H.~J.~McKellar, \prd{51} (1995) 2475; 
   H.~Anlauf, W.~Bernreuther, and A.~Brandenburg, 
   \prd{52} (1995) 3803 and references therein.
%
\bibitem{CR} F.~Cuypers and S.D.~Rindani, \plb{343} (1995) 333. 
%
\bibitem{Chang} D.~Chang, W.~Y.~Keung, and I.~Phillips, 
   \npb{408} (1993) 286; {\bf 429} (1994) 255 (E).
%
\bibitem{Atwood} D.~Atwood and A.~Soni, \prd{45} (1992) 2405.
%
\bibitem{PR} P.~Poulose and S.D.~Rindani, \plb{349} (1995) 379.
%
\bibitem{HPZH} K.~Hagiwara, R.D.~Peccei, D.~Zeppenfeld and 
   K.~Hikasa, \npb{282} (1987) 253.
%
\bibitem{GgDs} G.~Gounaris, D.~Schildknecht, and F.M.~Renard, 
   \plb{263}, 291 (1991); M.B.~Gavela, F.~Iddir, A.~Le
   Yaouanc, L.~Oliver, O.~P\'{e}ne, and J.C.~Raynal, 
   \prd{39}, 1870 (1989); A.~Bilal, E.~Mass\'{o},
   and A.~De R\'{u}jula, \npb{355}, 549 (1991).
%
\bibitem{PkPm} P.~Kalyniak, P.Madsen, N.~Sinha, and R.~Sinha,
   \prd{52}, 3826 (1995);
   D.~Chang, W.~Y.~ Keung, and I.~Phillips, \prd{48} (1993) 4045; 
   G.~Gounaris, D.~Schildknecht, and F.~M.~Renard, \plb{263} (1991) 291;
   M.~B.~Gavela, F.~Iddir, A.~Le Yaouanc, L.~Oliver, O.~P\'ene, and 
   J.~C.~Raynal, \prd{39} (1989) 1870; 
   A.~Bilal, E.~Mass\'o, and A.~De R\'ujula, \npb{355} (1991) 549.
%
\bibitem{Gounaris} G.J.~Gounaris and G.P.~Tsirgoti,
   Preprint No. THES-TP 97/03, E-preprint No. hep-ph/9703446.
%
\bibitem{Atwood2} D.~Atwood, G.~Eilam, and A.~Soni, \prl{71} 
   (1993) 492;
   B.Grzadkowski and J.~F.~Gunion, \plb{287} (1992) 237.
%
\bibitem{Wudka} J.~Wudka, Nucl.~Phys.~Proc.~Suppl.~37A (1994) 211.
%
\bibitem{GbGc} G.~B\'{e}langer and G.~Couture, 
   \prd{49} (1994) 5720.
%
\bibitem{KLY} G.L.~Kane, G.A.~Ladinsky, and C.-P.~Yuan,
   \prd{45} (1992) 124; C.R.~Schmidt and M.E.~Peskin, \prl{69} (1992) 410;
   D.~Atwood, A.~Aeppli, and A.~Soni, \prl{69} (1992) 2754.
%
\bibitem{Hikasa} K.~Hikasa, \plb{143}, 266 (1984);
   \prd{33}, 3203 (1986).
%
\bibitem{Kramer} M.~Kr\"{a}mer, J.~K\"{u}hn, M.L.~Stong and
   P.M.~Zerwas, \zpc{64} (1994) 21.
%
\bibitem{Model} W.~Buchm\"{u}ller and D.~Wyler, 
   \npb{268} (1986) 621;
   C.~J.~C.~Burges and H.~J.~Schnitzer, {\it ibid} {\bf B228} (1983) 464;
   C.~N.~Leung, S.~T.~Love and S.~Rao, \zpc{31} (1986) 433;
   K.~Hagiwara, S.~Ishihara, 
   R.~Szalapski, D.~Zeppenfeld, \prd{48} (1993) 2182; 
   K.~Hagiwara and M.~Stong, \zpc{62} (1994) 99.
%
\bibitem{BN} W.~Bernreuther and O.~Nachtmann, \prl{63} 
   (1989) 2787; W.~Bernreuther, O.~Nachtmann, P.~Overmann and 
   T.~Schr\"{o}der, \npb{388} (1992) 53. 
%
\bibitem{Top} F.~Abe {\it et al.} (CDF Collaboration),
   \prl{74} (1995) 2626;
   S.~Abachi {\it et al.} (D0 Collaboration), {\it ibid.} {\bf 74}
   (1995) 2632.
%
\bibitem{PDG96} Particle Data Group, L.~Montanet {\it et al.},
   \prd{54} (1996) 94.
%
\bibitem{BDKKZ} Ikaros I.~Y.~Bigi, Yu L.~Dokshitzer, V.~Khoze, 
   J.~K\"{u}hn and P.~Zerwas, \plb{181} (1986) 157.
%
\bibitem{Barr} S.M.~Barr and W.J.~Marciano, in {\it CP Violation},
   ed. C.~Jarlskog (World Scientific, Singapore, 1990).
%
\bibitem{Soni-Xu} A.~Soni and R.M.~Xu, \prl{69} (1992) 33.
%
\bibitem{CJK} M.~Jezabek and J.H.~K\"{u}hn, \npb{320} (1989) 20; 
   A.~Czarnecki, M.~Jezabek and J.H.~K\"{u}hn, \npb{351} (1991) 70.
%
\bibitem{Tsai} Y.-S.~Tsai, \prd{4} (1971) 2821;
   S.Y.~Pi and A.I.~Sanda, Ann.~Phys.~(N.Y.) {\bf 106} (1977) 16;
   H.~K\"{u}hn and F.~Wagner, \npb{236} (1984) 16;
   K.~Hagiwara, A.D.~Martin and D.~Zeppendeld, \plb{235}
   (1989) 198; B.K.~Bullock, K.~Hagiwara and A.D.~Martin, 
   \prl{67} (1991) 3055; \npb{395} (1993) 499;
   P.~Privitera, \plb{308} (1993) 163.
%
\bibitem{Grossman} Y.~Grossman, \npb{426} 
   (1994) 355 and references therein.
%
\bibitem{Falk} D.~Atwood, G.~Eilam and A.~Soni, 
   \prl{71} (1993) 492; A.F.~Falk, Z.~Ligeti, M.~Neubert 
   and Y.~Nir, \plb{326} (1994) 145; 
   Y.~Grossman and Z.~Ligeti, {\it ibid.} {\bf 332}, (1994) 373.
%
\bibitem{Nachtmann1} W.~Bernreuther, O.~Nachtmann and P.~Overmann,
   \zpc{48} (1993) 78 and references therein.
%
\bibitem{Kilan} U.~Kilan, J.G.~K\"{o}rner, K.~Schilcher and W.L.~Wu,
   \zpc{62} (1994) 413.
%
\bibitem{Rindani} B.~Ananthanarayan and S.D.~Rindani, \prl{73}
   (1994) 1215.
%
\bibitem{Tsai1} Y.S.~Tsai, \plb{378} (1996) 272 and references therein.
%
\bibitem{Huang} T.~Huang, W.~Lu and Z.~Tao, \prd{55} (1997) 1643.
%
\bibitem{Mirkes} J.H.~K\"{u}hn and E.~Mirkes, \plb{398} (1997) 407.
%
\bibitem{OPAL} R.~Akers {\it et al.} (OPAL Collaboration), 
    \zpc{66} (1995) 31.
%
\bibitem{ALEPH} D.~Buskulic {\it et al.} (ALEPH Collaboration), 
   \plb{346} (1995) 371.
%
\bibitem{Delaguila} F.~del Aguila and M.~Sher, \plb{252} 
   (1990) 116.
%
\bibitem{Grifols} J.A.~Grifols and A.~Mendez, 
   \plb{255} (1991) 611.
%
\bibitem{Escribano} R.~Escribano and E.~Masso, \plb{301} 
   (1993) 419.
%
\bibitem{Nachtmann} O.~Nachtmann, private communication.
%
\bibitem{Hanai} H.~Hanai {\it et al.} (Venus Collaboration),
   Preprint No. KEK preprint 96-171 (1996).
%
\bibitem{CLEOII} J.P.~Alexander {\it et al.} (CLEO Collaboration),
   Preprint No. CLNS 97/1480, CLEO 97-9.
%
\bibitem{Choi4} M.S.~Baek, S.Y.~Choi and K.~Hagiwara, 
   in preparation.
%
\bibitem{Jadach} S.~Jadach, J.H.~K\"{u}hn and Z.~Was, 
   Comput.~Phys.~Commun, {\bf 64} (1991) 275; M.~Jezabek, Z.~Was, S.~Jadach,
   and J.H.~K\"{u}hn, {\it ibid.} {\bf 70} (1992) 69;
   S.~Jadach, Z.~Was, R.~Decker, and J.H.~K\"{u}hn, {\it ibid.}
   {\bf 76} (1993) 361.
%
\bibitem{Tornqvist} N.A.~Tornqvist, \zpc{\bf C36} 
   (1987) 695; {\bf 40} (1988) 632; M.G.~Bowler, \plb{209} 
   (1988) 99; Yu.P.~Ivanov, A.A.~Osipov and M.K.~Volkov, \zpc{49} 
   (1991) 563.
%
\bibitem{Isgur} N.~Isgur, C.~Morningstar and C.~Reader, 
   \prd{39} (1989) 1357.
%
\bibitem{Decker} R.~Decker, E.~Mirkes, R.~Sauer and Z.~Was,
   \zpc{58} (1993) 445.
%
\bibitem{Kuhn} J.H.~K\"{u}hn, \plb{313} (1993) 458.
%
\bibitem{Davies} A.J.~Davies and X.-G.~He, \prd{43}
   (1991) 225.
%
\bibitem{Hall} L.J.~Hall and L.J.~Randall, \npb{274}
   (1986) 157; J.F.~Nievies, {\it ibid.} {\bf B189} (1981) 382;
   W.~Buchm\"{u}ller, R.~R\"{u}ckl and D.~Wyler, \plb{191}
   (1987) 44.
%
\bibitem{Dohmen} C.~Dohmen {et al.}, \plb{317}
   (1993) 631.
%
\bibitem{Marciano} W.J.~Marciano and A.~Sirlin, 
   \prl{71} (1994) 3629.
%
%
\bibitem{WEDM} W.~Marciano and A.~Queijeiro, \prd{33},
   3449 (1986); F.~Boudjema, K.~Hagiwara, C.~Hamzaoui, and
   K.~Numata, {\it ibid} D {\bf 43}, 2223 (1991);
   A.~De R\"{u}jula, M.~Gavela, O.~P\'{e}ne, and F.~Vegas,
   \npb{357}, 311 (1991); see also S.~Barr and 
   W.~Marciano, {\it CP Violation}, ed. C.~Jarlskog 
   (World Scientific, Singapore, 1989); D.~Chang, W.Y.~Keung and J.~Liu, 
   \npb{355} (1991) 295; R.~Lop\'{e}z-Mobilia and T.H.~West, 
   \prd{51} (1995) 6495.
%
\end{thebibliography}
\end{document}